\documentclass[11pt]{aa}
\usepackage[english]{babel}
\usepackage{graphicx}
\usepackage{longtable}

\onecolumn
\newcommand{\ab}{Astrophys. Bull. }
\newcommand{\arep}{Astron. Rep. }
\newcommand{\alet}{Astron. Let. }
\newcommand{\araa}{Ann. Rev. Astron. Astrophys. }
\newcommand{\mnras}{Mon. Not. R. Astron. Soc. }
\newcommand{\apj}{Astrophys. J. }

\newcommand{\aj}{Astron. J. }
\newcommand{\aaa}{Astron and Astrophys.}
\newcommand{\aas}{Astron and Astrophys. Suppl.}

\begin{document}

\title{Spectral atlas of  A--type supergiants}

\author{V.G.~Klochkova\thanks{E-mail: valenta@sao.ru}, E.G.~Sendzikas,  E.L.~Chentsov}
\institute{Special Astrophysical Observatory RAS, Nizhnij Arkhyz,  369167 Russia}

\date{\today}
\titlerunning{\it Spectral atlas of  A--type supergiants}

\authorrunning{\it Klochkova et al.}

\abstract{Based on high--spectral--resolution observations  ($R=60\,000$) performed with 
  the 6-m BTA telescope in combination with the echelle spectrograph NES, we have studied 
  the optical spectra of three A--type supergiants: a peculiar supergiant 3\,Pup, a post-AGB star 
  BD\,$+48\degr1220$, and a massive $\alpha$\,Cyg, which belong to essentially different stages of evolution. 
  A spectral atlas for these stars is prepared  in the wavelength interval of 3920 to 6720\,\AA{}.
}

\maketitle

\section{Introduction}

The present paper is the next step in the study of evolved high luminosity stars (HLSs) 
at short final phases when a star is still powered by nuclear fusion. Our objects of study are
both the most massive stars with initial masses greater than 10~$M_{\odot}$ presumably at the 
stages of LBV, B[e] supergiants, yellow hypergiants (YHGs), the luminosity of which is often close
to the Eddington limit, and the medium mass stars with initial masses of 3--8~$M_{\odot}$ at the 
asymptotic giant branch (AGB) phase and the further post-AGB phase. The identification of stars
of these types (LBV, B[e], YHG, AGB, or post--AGB), reliable determination of the fundamental parameters, 
and defining the evolutionary stage of HLSs have become particularly important recently owing to the 
search for stars of these types in the Local Group galaxies (see~[\cite{Massey, Hump2013}] and references therein). 

The high luminosity stars of the types mentioned above, which fundamentally differ in mass and evolutionary 
stage, have similar observable properties: the characteristics of the optical and radio spectra, large IR excesses, 
complex and variable velocity fields; this shows the instability of their extended atmospheres and expanding
gaseous-dusty envelope. This overlapping of the observed characteristics of two types of objects represents 
the so-called ``spectroscopic mimicry'' problem.

The main feature of the optical spectra of the high and low mass HLSs are the absorption and emission profiles 
of the spectral lines, sometimes of an anomalous form, primarily of H$\alpha$ lines. This can be a direct or an 
inverse P\,Cyg profile, an absorption with asymmetric emission wings, or a combination of different elements. 
We remind that the emission in H$\alpha$ along with the IR excess is a principal selection criterion for post-AGB
star candidates~[\cite{Kwok1993}]. The emission profile of H$\alpha$ is a sign of the outflow of matter and/or pulsations.
The differences in the profiles  reflect the differences in the dynamic processes which occur in the extended 
atmospheres of some stars: the spherically symmetric outflow, infall of matter onto the photosphere, pulsations.

The best known example of an undefined evolutionary stage is the case of a peculiar supergiant V1302\,Aql 
(IR-source IRC\,+10420), which had been considered a post-AGB stage representative for several decades until now. 
However, some authors referred this unique object to the most massive stars of the Galaxy, because some 30 years 
ago it was noted to be similar to $\eta$\,Car, the distinguished LBV of the Galaxy. The cumulative results obtained 
with long-term observations using various techniques at the world's largest telescopes, and the 6--m BTA telescope
among them~[\cite{IRC1,IRC2}], revealed the actual state of things. Now this object is conclusively classified as 
a massive star of  extreme luminosity---a yellow hypergiant.

The poorly studied variable A~supergiant  V2324\,Cyg identified with the IRAS\,20572+4919 IR~source can serve also 
as an example of an object with an undetermined status. The observable excess of radiation at about 12--60~$\mu$m and 
the position on the IR-color diagram allow us to refer this object to the post-AGB stage, i.e., to consider it a 
protoplanetary nebula (PPN) with a dust shell. Using the BTA spectra for V2324\,Cyg, with the method of model 
atmospheres we determined $T_{\rm eff}=7500$\,K, surface gravity  $\log g=2.0$, microturbulence velocity 
$V_t=6.0$\,km\,s$^{-1}$, and solar metallicity~[\cite{20572}]. An unexpected feature  for a PPN is lithium and 
sodium overabundance. The overabundance of these elements in the atmosphere allows us to classify V2324\,Cyg
as a star with the initial mass greater than 4~$M_{\odot}$. Nuclear fusion of light metals (Li,~Na,~Al) is 
possible owing to the hot bottom burning (HBB) in the hot layers of the convective envelope 
in massive AGB-stars. The description of HBB and the necessary references are available in Ventura et al.~[\cite{Ventura}].

Despite the large amount of obtained data, the evolutionary status of V2324\,Cyg is still unclear. In the IR-color diagram, 
the IRAS\,20572+4919 source is located in region IV, which is occupied by planetary and protoplanetary nebulae. In accordance 
with the chronological sequences presented by Lewis~[\cite{Lewis}], the absence of maser radiation in OH and H$_2$O lines indicates 
the approach of the object to the planetary nebula phase. This fact agrees with the conclusions from the publications in which 
a number of authors~[\cite{Garcia,Arkhip}] regard V2324\,Cyg as a post-AGB star. However, some characteristics of
the star do not correspond to the post-AGB stage; the first one is its low luminosity: the spectral classification indicates 
that its luminosity class is III. An excessive surface gravity, $\log g=2.0$~\cite{20572}, indicates the same. The H$\alpha$
line profile and the high wind velocity, typical for supergiants, do not correspond to the post-AGB star status as well. As follows
from Fig.\,1 in Klochkova and Chentsov~[\cite{20572}], the H$\alpha$ profile with a strong variable emission more likely
implies a high luminosity star with fast wind.

To gain experience and set up the classification criteria for HLSs, it would be useful to compare in detail their high 
resolution spectra in a wide wavelength interval. The present paper contains the atlas of the optical spectra of three 
supergiants similar in spectral type but different in luminosity: a massive supergiant $\alpha$\,Cyg, a peculiar supergiant 
3\,Pup (A2.7\,Ib), and a protoplanetary nebula BD\,$+48\degr1220$ (A4\,Ia). The star 3\,Pup 
(HD\,62623\,$=$\,HR\,2996\,$=$\,MWC\,570\,$=$\,HIP\,37677)  is listed as an A2\,Iabe supergiant in the Bright 
Star Catalogue~[\cite{Bright}].

After reviewing the main characteristics and parameters of the stars under study in Section~\ref{descrip}, we will briefly
present the used observational data (Section~\ref{obs}) and the compiled spectral atlas (Section~\ref{atlas}).

\begin{figure*}
\hbox{
\includegraphics[angle=0,width=0.3\textwidth,bb=20 40 340 530,clip]{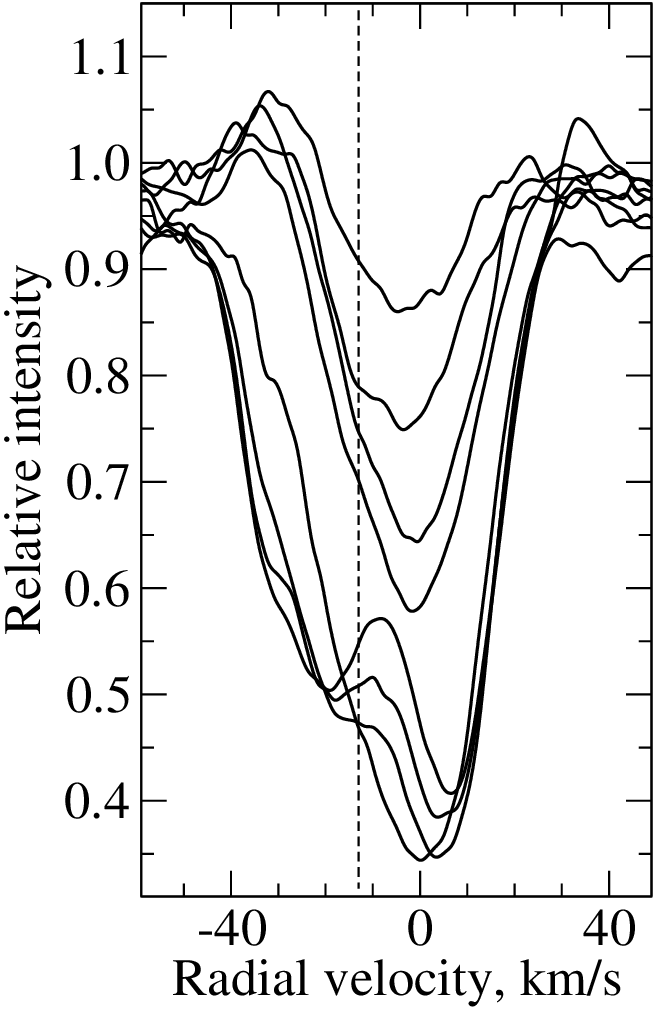}
\includegraphics[angle=0,width=0.3\textwidth,bb=20 40 340 530,clip]{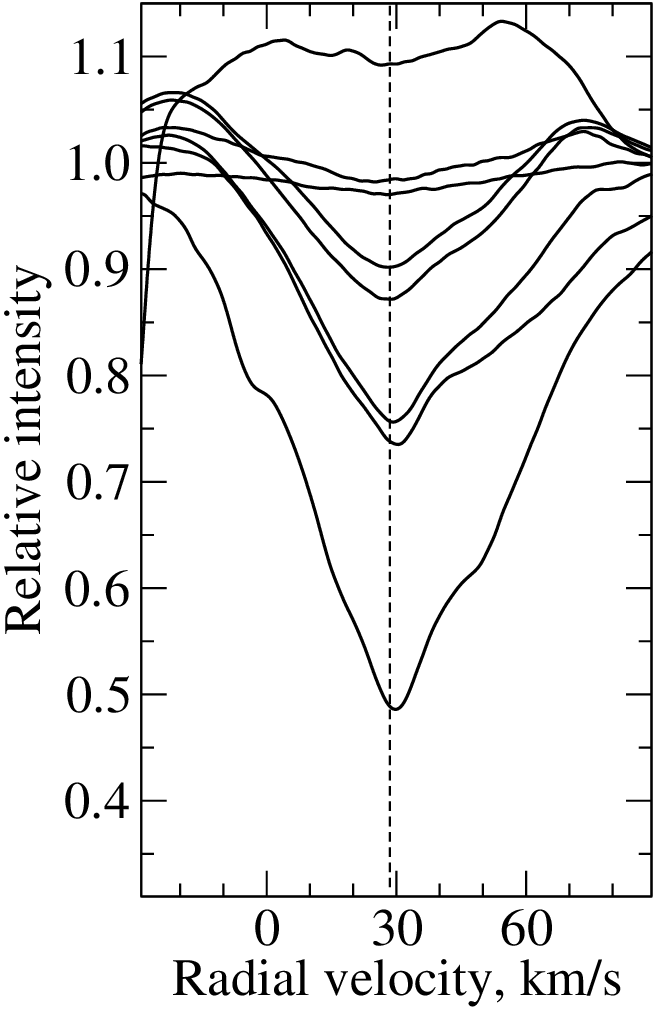}
\includegraphics[angle=0,width=0.3\textwidth,bb=20 40 340 530,clip]{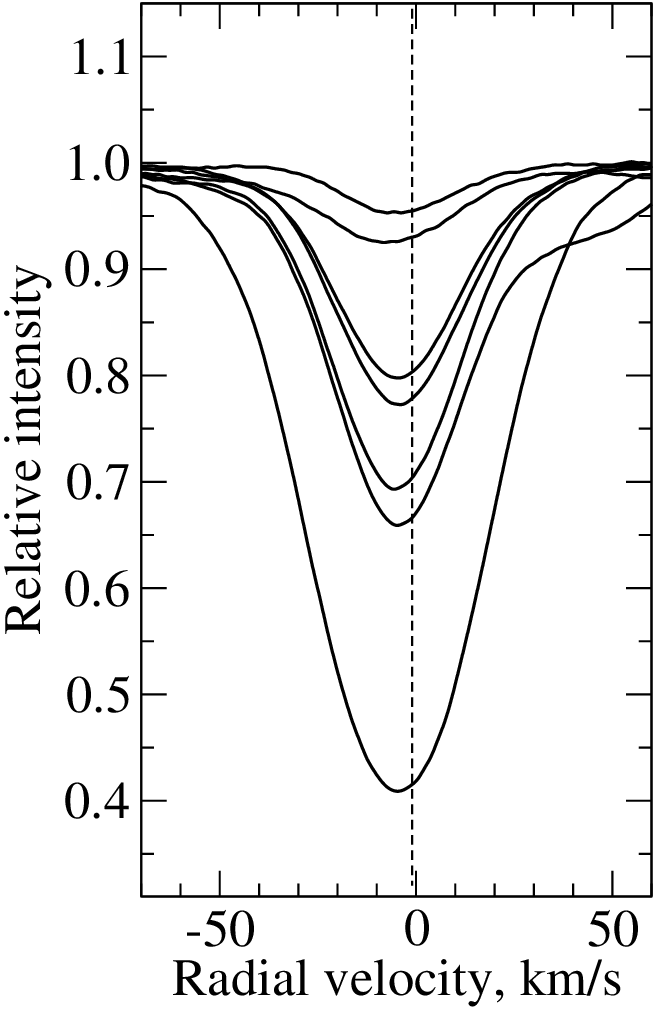}
}
\caption{Line profiles of different intensity in the spectra of the studied stars. Left to right:
         BD\,$+48\degr1220$ (top to bottom: Fe\,II 6433, 6248, 5535, 5363, 5169, 5018, 4924, 4523\,\AA{}), 
         3\,Pup (top to bottom: [O\,I] 6300\,\AA{}, Fe\,II 6369, 6084, 5284, 5535\,\AA{}; Ti\,II 4501,
         4444\,\AA{}; Fe\,II 4233\,\AA{}), and $\alpha$\,Cyg (top to bottom: Fe\,II 4433\,\AA{}; Ti\,II 
         4444, 4501\,\AA{}; Fe\,II 5535, 5284, 5015, 6084\,\AA{}). The vertical dashed line denotes the systemic
         velocity $V_{\rm sys}$ from~[\cite{05040b,3Pup,Kaufer}] for BD\,$+48\degr1220$, 3\,Pup, and 
         $\alpha$\,Cyg respectively.}
\label{Profiles}
\end{figure*}

\section{Characteristics of the studied A--type supergiants}\label{descrip}

The studied A supergiants are presented in Table~\ref{stars} in the same order as their spectra
are presented in the atlas. Along with the star's name, the table gives the numbers of the associated IR-sources, 
the spectral type according to previous investigations of the stars, their evolutionary status,
absolute magnitude $M_V$, visible magnitude~$V$, and galactic coordinates $l$, $b$. The last two columns of
this table show the mid-exposure time (in JD) and the heliocentric radial velocity $V_r$ measured by numerous 
symmetric absorptions in the spectra of the studied stars. Let us focus on the observed parameters and characteristics 
of the stars.

\begin{table*}
\caption{General characteristics of the studied stars. The second column shows the spectral class and type of supergiant. 
         Absolute $M_V$magnitudes for $\alpha$\,Cyg,  3\,Pup, and BD\,$+48\degr1220$ from~[\cite{Schiller}], [\cite{3Pup}], 
         and~[\cite{05040b}] respectively,  are given in the third column. The last two columns show the mid-exposure 
         time in JD and the heliocentric radial velocity  $V_r$ measured by numerous ($n> 500$) symmetric absorptions}
\medskip
\begin{tabular}{l|c|l|c|c|c|c|r@{$\,\pm\,$}l}
\hline
 \multicolumn{1}{c|}{Star}      & $Sp$,  & \multicolumn{1}{c|}{$M_V$,}   & $V$, & $l$, & $b$, & JD\,2456000 & \multicolumn{2}{c}{$V_r$,} \\
 \multicolumn{1}{c|}{IR-source} & status &  \multicolumn{1}{c|}{mag}     & mag  & deg  & deg  &             & \multicolumn{2}{c}{km\,s$^{-1}$}\\
 \hline
 BD\,$+48\degr1220$ & A4\,Ia     & $-5.0$  & 9.74 & 159.7 & +04.8    & 581.4 &  $-5.5$ & $0.1$\\
 IRAS\,05040+4820   & post-AGB   &         &      &       &          &       &\multicolumn{2}{c}{}\\
\hline
 3\,Pup             & A2.7\,Ib   & $-5.5$  & 3.93 & 244.4 & $-02.5$  & 579.6 & $+30.5$ & $0.3$\\
 IRAS\,$07418-2850$ & peculiar &         &      &       &          &       &\multicolumn{2}{c}{}\\
\hline
 $\alpha$\,Cyg     & A2\,Iae     & $-8.38$ & 1.25 & 084.3 & +02.0    & 524.5 & $-5.2$ & $0.1$\\
 IRC\,+50337       & massive   &         &      &       &          &       &\multicolumn{2}{c}{}\\
\hline
\end{tabular}
\label{stars}
\end{table*}

\subsection{$\alpha$\,Cyg}

On the Hertzsprung--Russell diagram, the massive A supergiant $\alpha$\,Cyg is close to the Humphreys--Davidson limit. I
ts high luminosity (luminosity class~Ia) manifests itself in large-scale macroturbulence~[\cite{Jager}], and the
variability of brightness and velocity field in the extended shell. $\alpha$\,Cyg is a prototype of the same type of hot
variable supergiants, for which a low amplitude variability of brightness and radial velocity is typical. All the absorptions
except H$\alpha$ are symmetric in the star's spectrum (see an example in Fig.\,\ref{Profiles}). Owing to its
brightness and convenient location in the northern sky, the star is well studied and described in numerous papers. 
The main characteristics of the $\alpha$\,Cyg spectrum and its variability are presented by Richardson et al.~[\cite{Richard}], 
who carried out a long-term spectral monitoring with a high spectral resolution in combination with Str\"omgren system 
photometry. However, despite the long-term observations of $\alpha$\,Cyg, the nature of its variability remains unclear 
as long as the estimates of the star's mass and its brightness variability periods are not definitely stated~[\cite{Richard,Saio}].

\subsection{3\,Pup}

3\,Pup was earlier considered as an object related to the stars evolving to a double nucleus planetary
nebula~[\cite{Jura}]. The reason for such a classification was its untypical for massive A supergiants
feature---an extended envelope around the star, the dust component of which appears in the IR excess~[\cite{Meilland}].
According to the IRAS~[\cite{Trams}] data, 3\,Pup is identified with the IR source IRAS\,07418$-$2850. Only after 
determining the chemical composition of its atmosphere to be similar to Solar~[\cite{Plets}], a strong argument
arose for 3\,Pup to be a massive A[e] supergiant. The A[e]--B[e] phenomenon manifests itself in the combination
of strong Balmer emission lines, and forbidden emissions [Fe\,II], [O\,I] with a considerable IR-flux excess associated
with the hot dust shell. All these features are by definition typical for massive B[e] supergiants~[\cite{Lamers}] and
are observed in the 3\,Pup system.

Massive stars can be observed at the blue supergiant stage, or at the stage of evolving from the main sequence, 
or while evolving after the red supergiant stage. According to the review~[\cite{Oudm}], supergiants with the B[e]
phenomenon in the optical spectrum can be viewed as objects evolving towards the red supergiant stage.

The inner gas shell influences the formation of peculiar double-peak emission details in the optical spectra of 3\,Pup,
which had been closely studied using the data on the spectral monitoring of this star carried out at the BTA during
1997--2008~[\cite{3Pup}]. Moreover, in that paper we updated the spectral class (A2.7\,$\pm 0.3$\,Ib) using the
equivalent widths and profiles of H$\delta$ and H$\gamma$; and from the O\,I~7774\,\AA{} triplet intensity, we estimated
the luminosity $M_V=-5\fm5\pm 0\fm3$.

As is evident from Table~\ref{Lines}, which shows the depth and position measurements of the absorption and emission
line components, the shell influences the majority of the lines; moreover, this influence gradually increases with 
the growth of line intensity and wavelength.

Regardless of numerous investigations of this star, which is an MK classification standard, the question of its 
binarity still remains open~[\cite{Loden, Plets}]. By the faintest photospheric absorptions, Chentsov et
al.~[\cite{3Pup}] found considerable (up to 7\,km\,s$^{-1}$) velocity changes from date to date, which indicate
a companion star. The results of the monitoring by Chentsov et al.~[\cite{3Pup}] are in overall agreement with the
interpretation of Plets et al.~[\cite{Plets}] about 3\,Pup being a star with an equatorial disk and a low mass
companion.

Important data was obtained by Millour~et~al.~[\cite{Millour}] with the AMBER VLTI spectro-interferometer with millisecond 
spatial resolution, which allowed them to detect the dust and gas disks around 3\,Pup. The second companion was not found; 
however, owing to the high spectral resolution of the equipment, the authors studied the velocity field of the object and 
concluded that there should be a second component which can form the Kepler disk and transfer the angular momentum to it.

\subsection{BD\,$+48\degr1220$}

The supergiant BD\,$+48\degr1220$ (SAO\,40039, LSV$+48\degr26$) is an optical component of the  IRAS\,05040+4820 IR source. 
The star is rather close to the Galactic plane ($b = 4\fdg8$), which can indicate its possible membership in the galactic 
disk population. Having multicolor photometric observations in the optical range for a sample of high luminosity stars, 
Fujii et al.~[\cite{Fujii}] obtained for BD\,$+48\degr1220$ the stellar magnitudes $B=10\fm1$,  $V=9\fm65$ and the A4\,Ia 
spectral class. The high luminosity and the double-humped distribution of energy in the spectrum, indicating the presence 
of circumstellar matter which was ejected in the course of previous evolution, allow us to include BD\,$+48\degr1220$ 
in the post-AGB star group.

Based on high spectral resolution ($R$\,=\,60\,000) observations, Klochkova et al.~[\cite{05040a}] found the
variability of the profiles and line positions. Afterwards, Klochkova et al.~[\cite{05040b}] thoroughly studied the
optical spectrum of BD\,$+48\degr1220$ in the wavelength range of 4500 to 6760\,\AA{} and determined the basic parameters 
of the star and the chemical composition of its atmosphere. With the method of model atmospheres, they obtained 
the following parameters: effective temperature $T_{\rm eff}$\,=\,7900\,K, surface gravity $\log g$\,=\,0.0, and 
microturbulence velocity  $\xi_t$\,=\,6.0\,km\,s$^{-1}$. The metallicity of the star is almost similar to the solar one: 
${\rm [Fe/H]}_{\odot}$\,=$-0.10$. However, its chemical composition contains a set of peculiarities: a large
helium abundance obtained from the He\,I\,$\lambda$~5876\,\AA{} absorption: ${\rm [He/H]}_{\odot}$\,=+1.04 along 
with a considerable oxygen abundance ${\rm [O/Fe]}_{\odot}$\,=+0.72. Also  the modified abundances of
light  metals: ${\rm [Si/Fe]}_{\odot}$\,=+0.81, ${\rm [Na/Fe]}_{\odot}$\,=+0.87 with  ${\rm [Mg/Fe]}_{\odot}$\,=$-0.31$ 
and the ratio ${\rm [Na/Mg]}_{\odot}$\,=+1.18 were found. A number of chemical composition
peculiarities (a close-to-solar metallicity, carbon, lithium, and sodium overabundances) allow us to consider
BD\,$+48\degr1220$ as an analogue of the peculiar A supergiant V2324\,Cyg, mentioned in the Introduction. 
The barium abundance in the BD\,$+48\degr1220$ atmosphere, however, is rather low: ${\rm [Ba/Fe]}_{\odot}$\,=$-0.84$, 
whereas in the case of V2324\,Cyg, one can observe its slight overabundance: ${\rm [Ba/Fe]}_{\odot}$\,=+0.46.

The luminosity of the star and the peculiarities of its chemical composition (helium, lithium, silicon, sodium, and oxygen
overabundances when ${\rm [C/O]}< 1$) allow us to refer BD\,$+48\degr1220$ to the more massive post--AGB stars, 
which have ongoing HBB nuclear reactions at the base of the hot convective shell at the AGB stage. Moreover,
we found distortions in the occurrence of chemical elements due to their selective separation into dust particles in the
circumstellar envelope. The argument for the separation is a high zinc abundance, $\rm {[Zn/Fe]}$\,=+0.44, and close values of
the calcium and scandium deficiency. The above-mentioned peculiarities of the abundance of chemical elements in the
atmosphere of BD\,$+48\degr1220$ were later confirmed by the echelle spectra from the McDonald and Vainu
Bappu~[\cite{Rao}] observatories. Altogether, the parameters ($M_V \approx -5^{\rm m}$, $V{\rm _{lsr} \approx
-20}$~km\,s$^{-1}$, metallicity ${\rm [Fe/H]}_{\odot}$\,=$-0.10$, peculiarities of the optical spectrum and chemical abundance)
confirm for BD\,$+48\degr1220$ the status of a He- and O-rich post-AGB star in the disk of the Galaxy with an initial
mass of $M>$4$M_{\odot}$. Note a rarely occurring feature of the chemical composition in the atmosphere of
BD\,$+48\degr1220$---a reliably estimated nickel overabundance.

Thus, for producing the atlas, we chose three A~supergiants with different masses and evolutionary stages. 
The observed characteristics of these stars differ fundamentally, which ensures the significance of comparing their optical spectra.

\section{Spectral data}\label{obs}

The spectra of all the objects that we used for producing the atlas were obtained in the Nasmyth focus of the 6-m telescope. The
echelle spectrograph NES~[\cite{nes}] in combination with a CCD of $2048\times4608$~elements and an image slicer~[\cite{slicer}] 
provides the spectral resolution $R=60\,000$. The Julian date for the mid-exposure time is given in the next-to-last column of
Table~\ref{stars}.

The extraction of data from two-dimensional echelle images was conducted with the modified~[\cite{ECHELLE}] ECHELLE context 
of ESO--MIDAS in the Linux operating system. Cosmic ray hit were removed by median averaging of two spectra obtained 
successively one after another. Wavelength calibration was done using the Th-Ar lamp with a hollow cathode. Further 
reduction of one-dimensional spectra, including continuum normalization, photometric and position measurements,
was performed using the computer code  DECH\,20T~[\cite{gala}] version 7.3.23. To increase the measurement accuracy 
of the intensity and spectral line positions by comparing the observed spectra with the corresponding synthetic
ones, unblended lines were selected. The zero-point position of every spectrogram was determined with the standard
technique---matching with the positions of telluric absorptions and ionospheric emissions which can be observed along 
with the spectrum of the object. The accuracy of single-line velocity measurements in the spectra obtained by NES is 
higher than 1\,km\,s$^{-1}$~[\cite{05040b}].

\section{Atlas description}\label{atlas}

The production of spectral atlases is particularly important in connection with the improvement in the quality of observational
data due to the use of up-to-date high resolution spectrographs that record spectra on low-noise CCDs. Furthermore, at present
there is a possibility of presenting the graphical information from the atlases and the detailed tables with line identifications
digitally, which provides users with unlimited access to this observational data.

\begin{figure*}
\vbox{
\includegraphics[angle=0,width=1.0\textwidth,bb=35 210 710 521,clip]{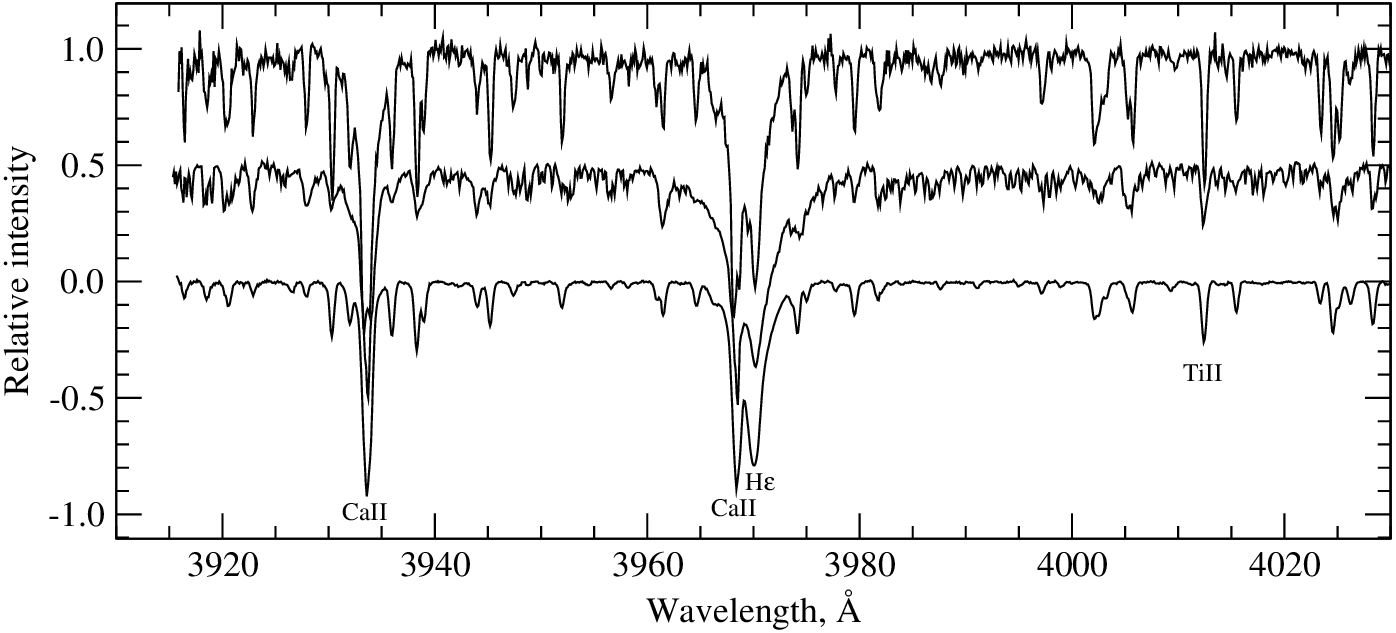}  
\includegraphics[angle=0,width=1.0\textwidth,bb=35 210 710 521,clip]{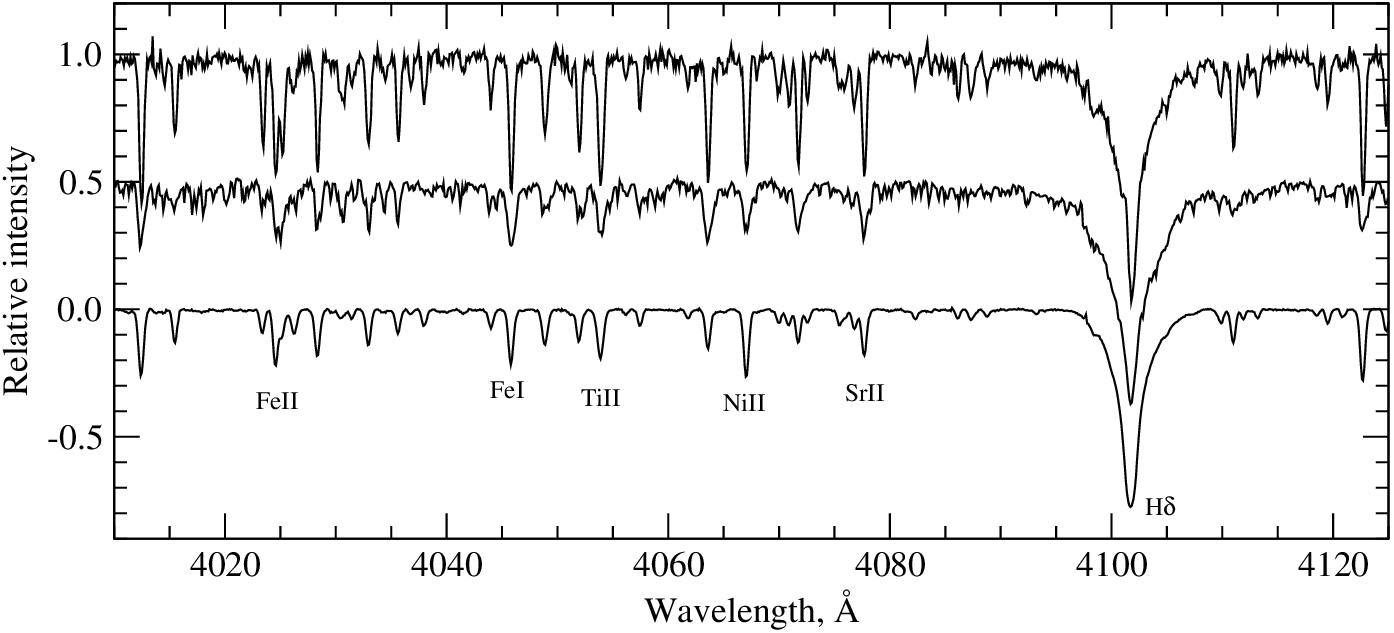} 
\includegraphics[angle=0,width=1.0\textwidth,bb=35 210 710 521,clip]{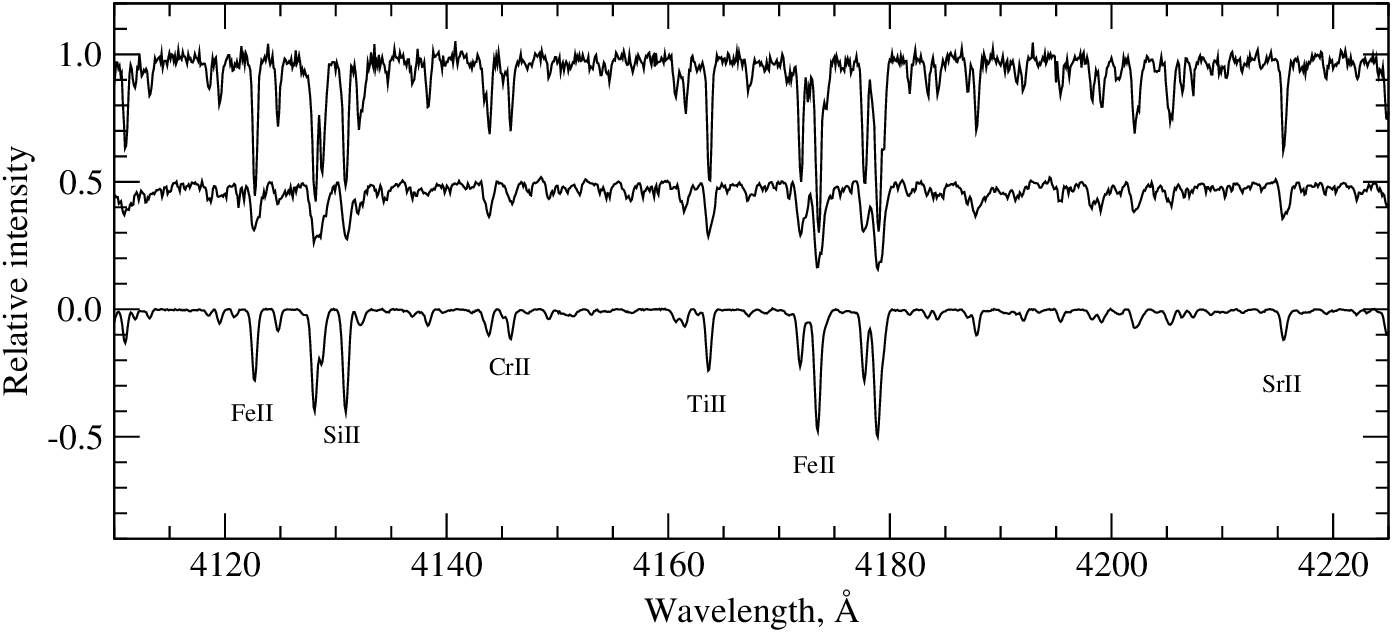}  
}
 \caption{Three shortwave fragments from the atlas in the wavelength range of 3920--4220~\AA{}. 
         Hereafter, the spectrum fragments are arranged top to bottom: BD$+48\degr1220$, 3\,Pup, $\alpha$\,Cyg,
         shifted along the vertical axis relative to each other.}
\label{fragment1}
\end{figure*}

\begin{figure*}
\vbox{
\includegraphics[angle=0,width=1.0\textwidth,bb=35 210 710 521,clip]{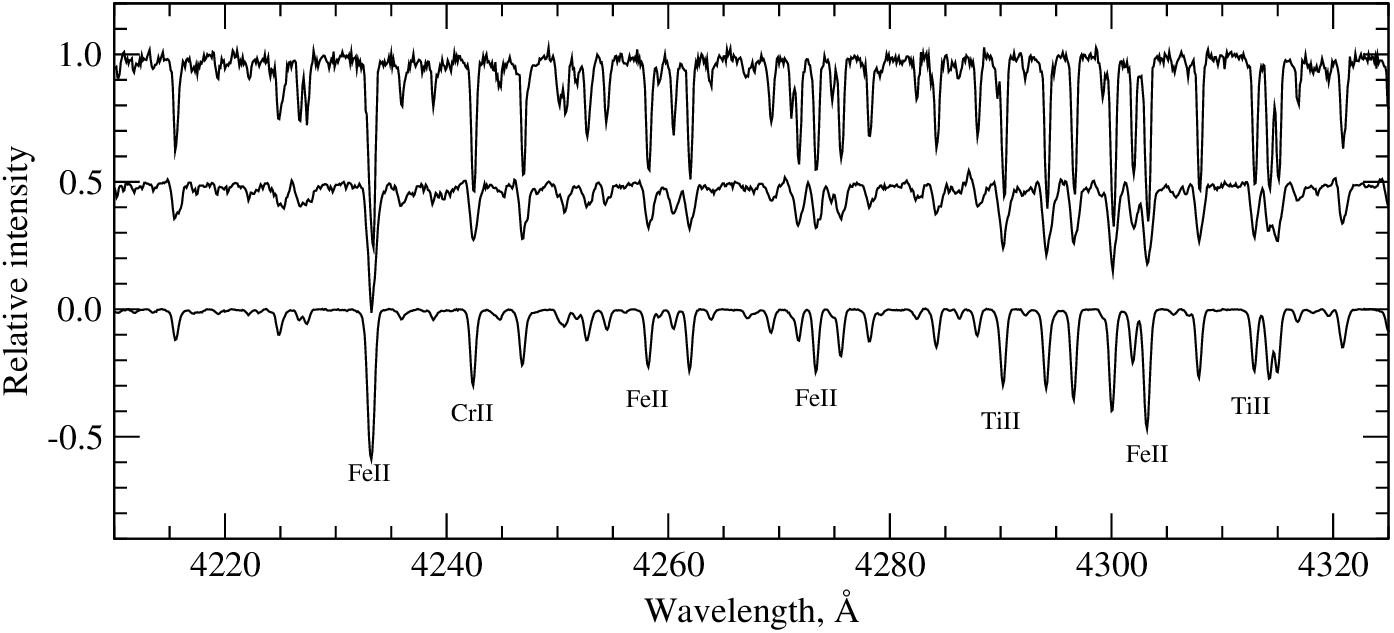}
\includegraphics[angle=0,width=1.0\textwidth,bb=35 210 710 521,clip]{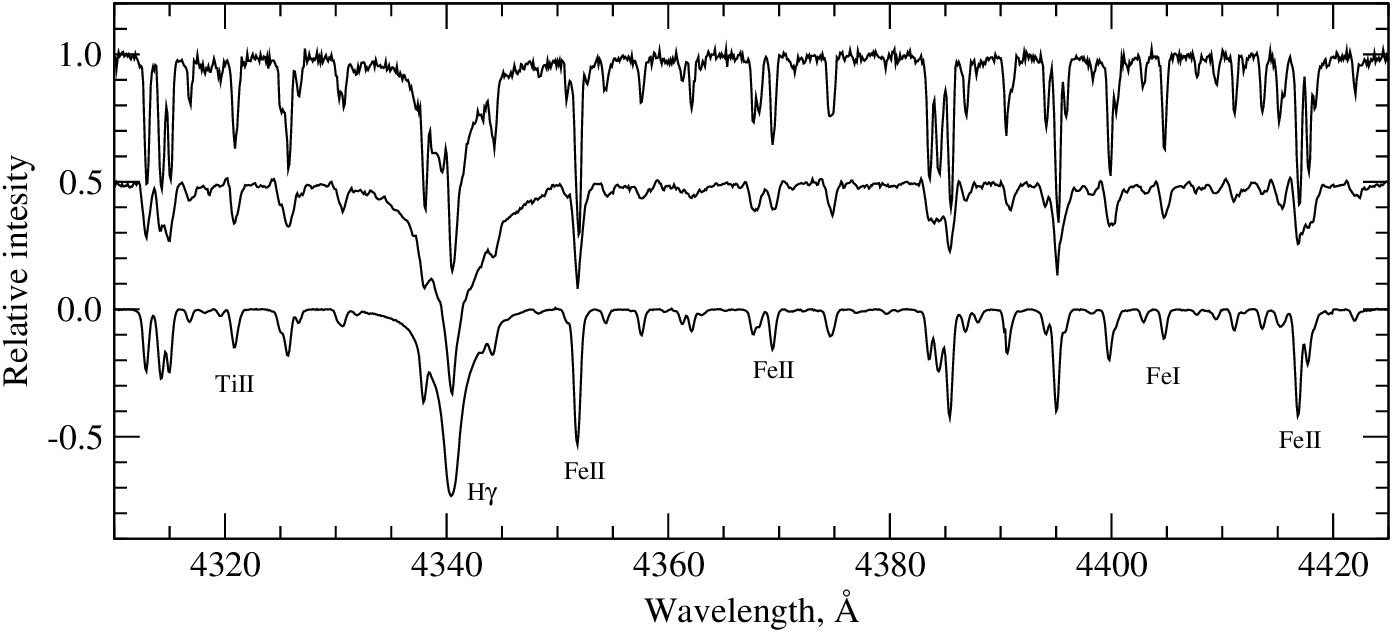}
\includegraphics[angle=0,width=1.0\textwidth,bb=35 210 710 521,clip]{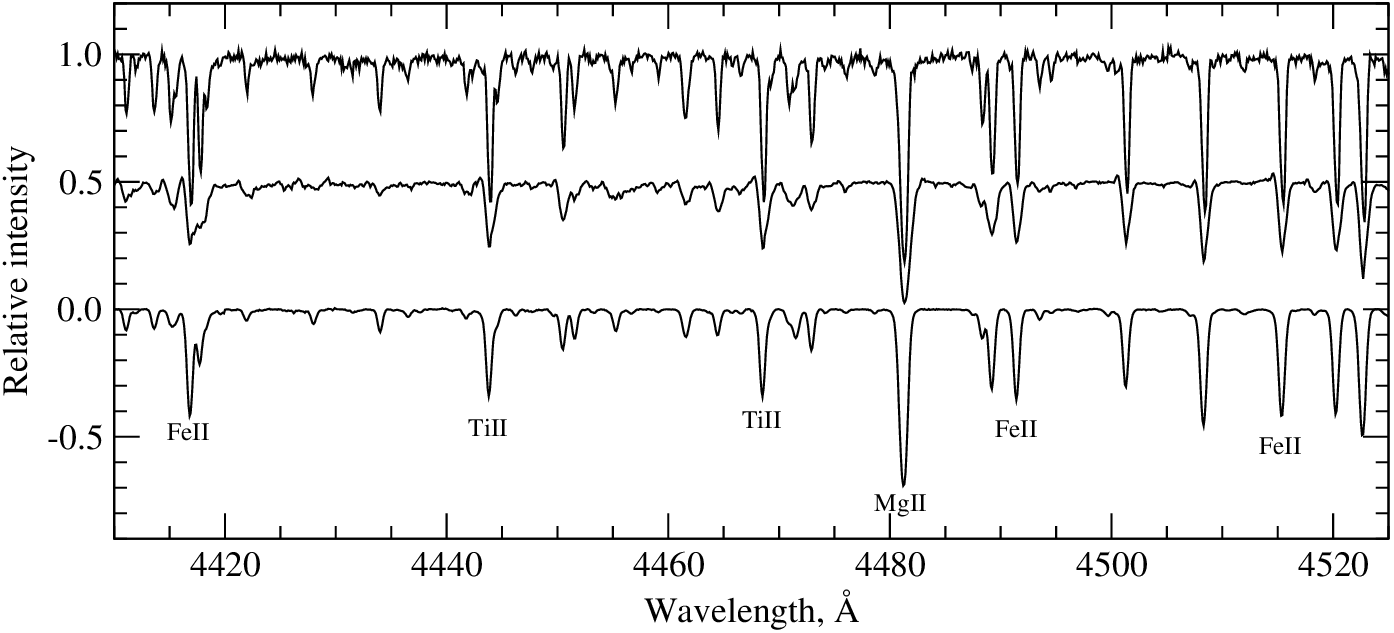}
}
\caption{} 
\label{fragment2}
\end{figure*}

\begin{figure*}
\vbox{
\includegraphics[angle=0,width=1.0\textwidth,bb=35 210 710 521,clip]{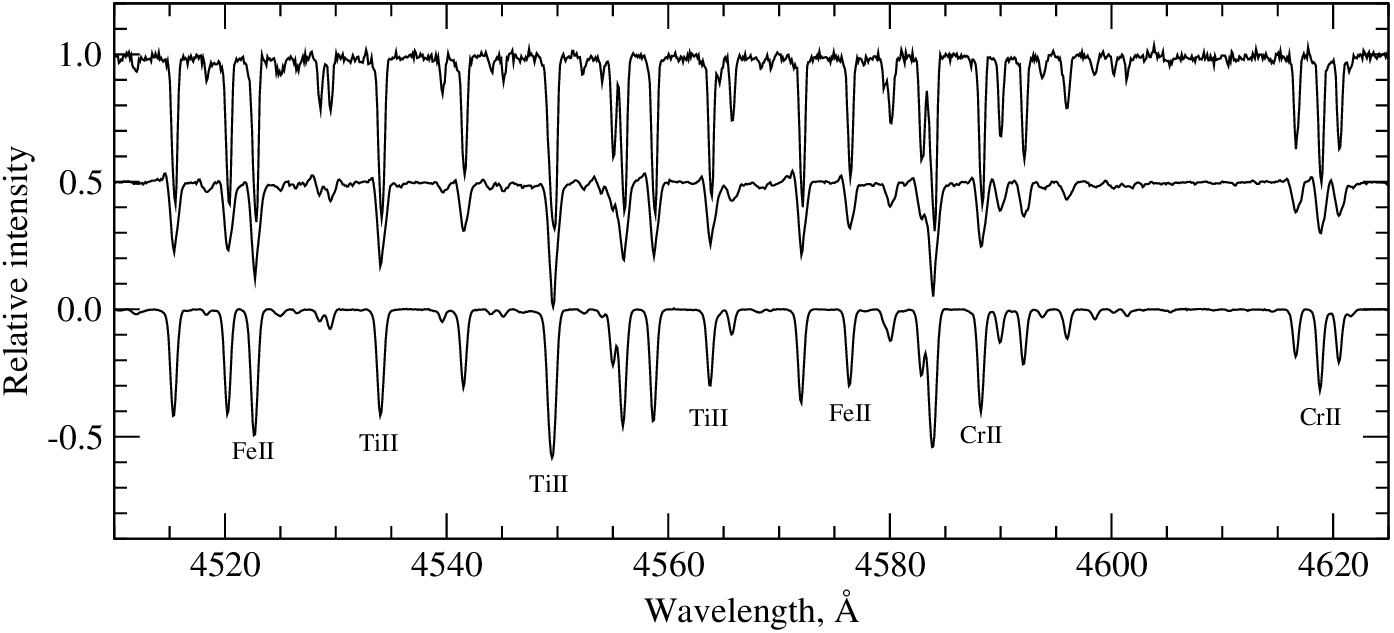}
\includegraphics[angle=0,width=1.0\textwidth,bb=35 210 710 521,clip]{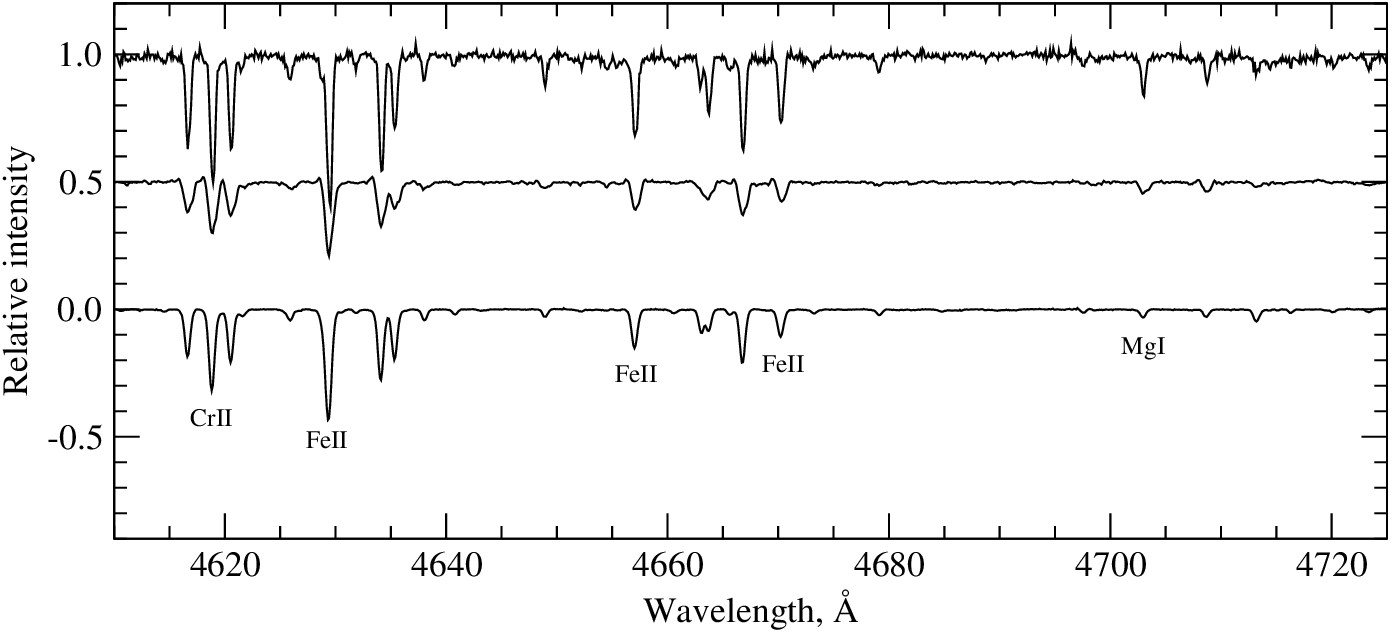}
\includegraphics[angle=0,width=1.0\textwidth,bb=35 210 710 521,clip]{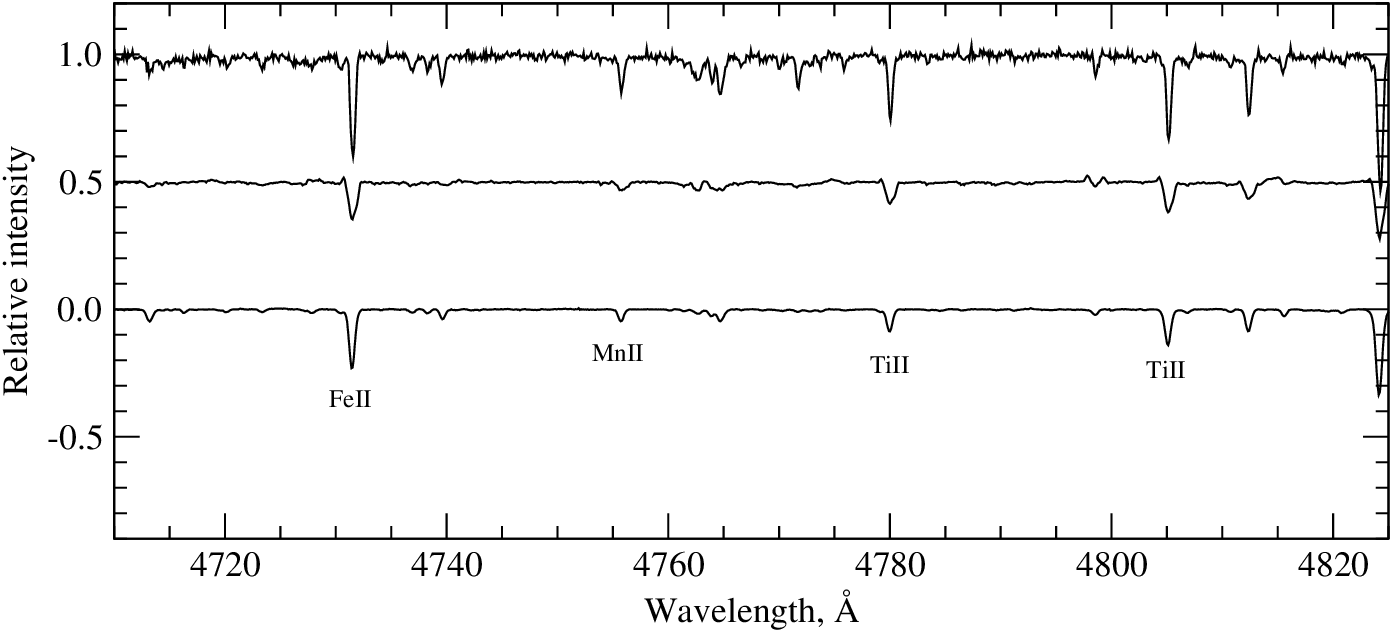}
}
\caption{} 
\label{fragment3}
\end{figure*}

\begin{figure*}
\vbox{
\includegraphics[angle=0,width=1.0\textwidth,bb=35 210 710 521,clip]{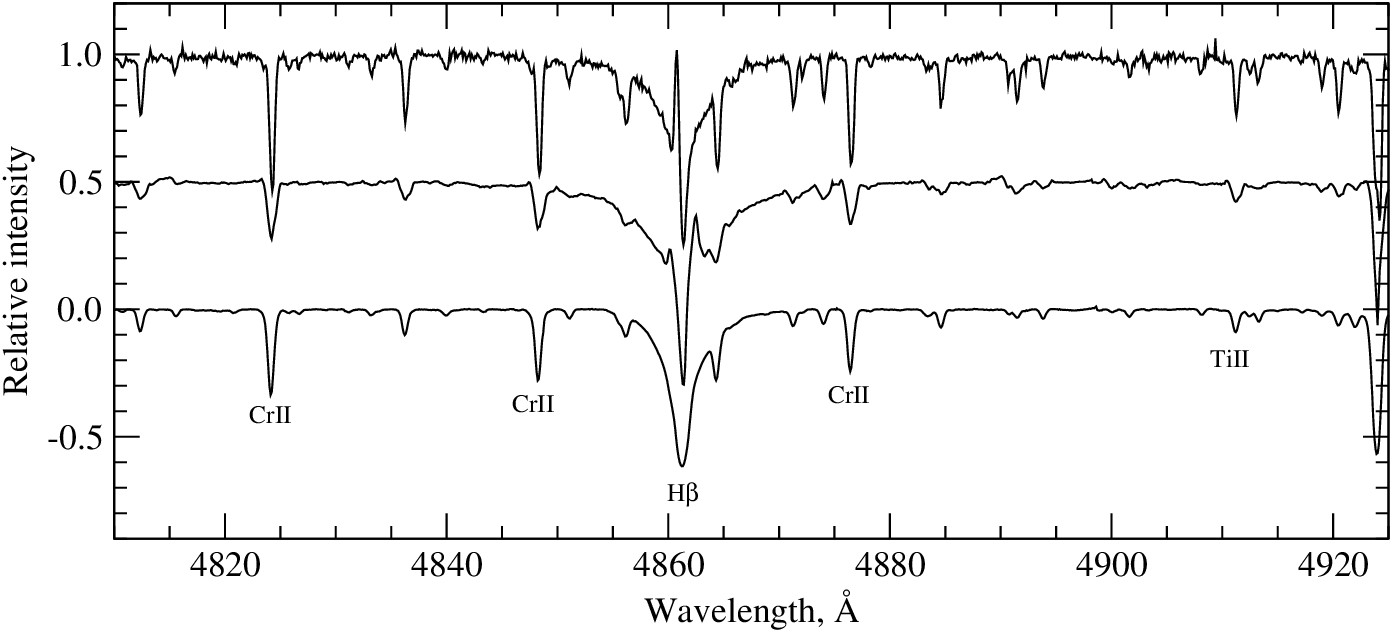}
\includegraphics[angle=0,width=1.0\textwidth,bb=35 210 710 521,clip]{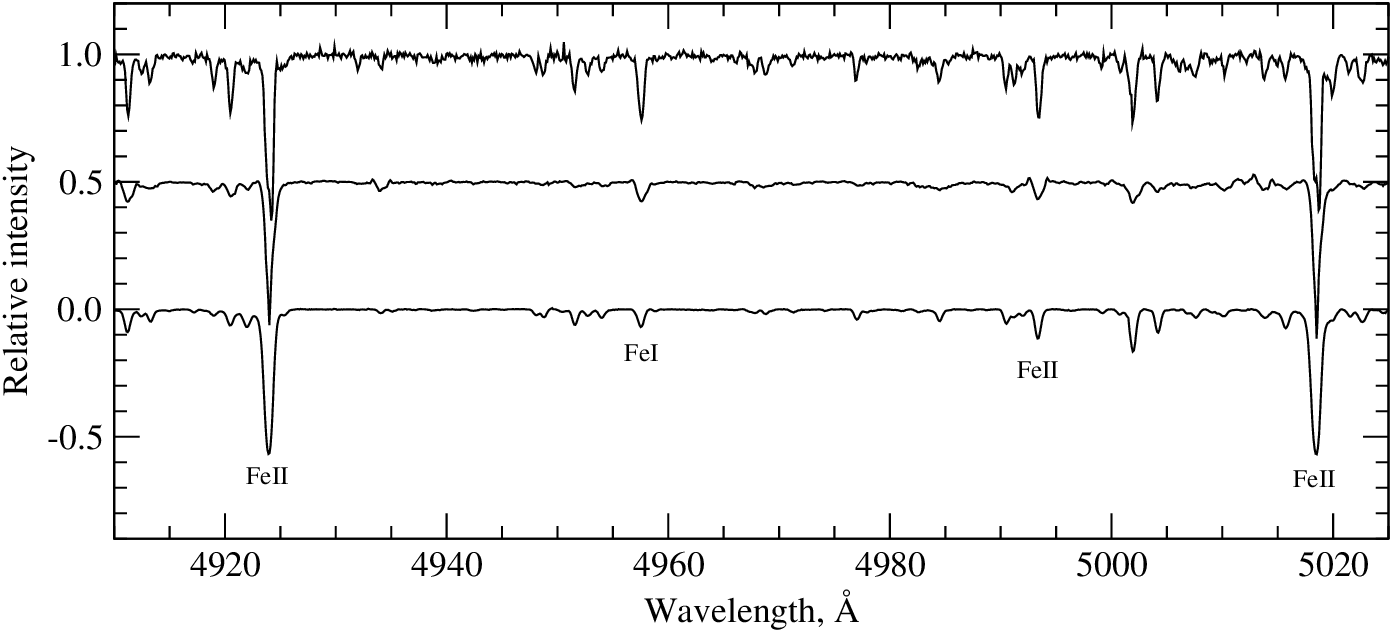}
\includegraphics[angle=0,width=1.0\textwidth,bb=35 210 710 521,clip]{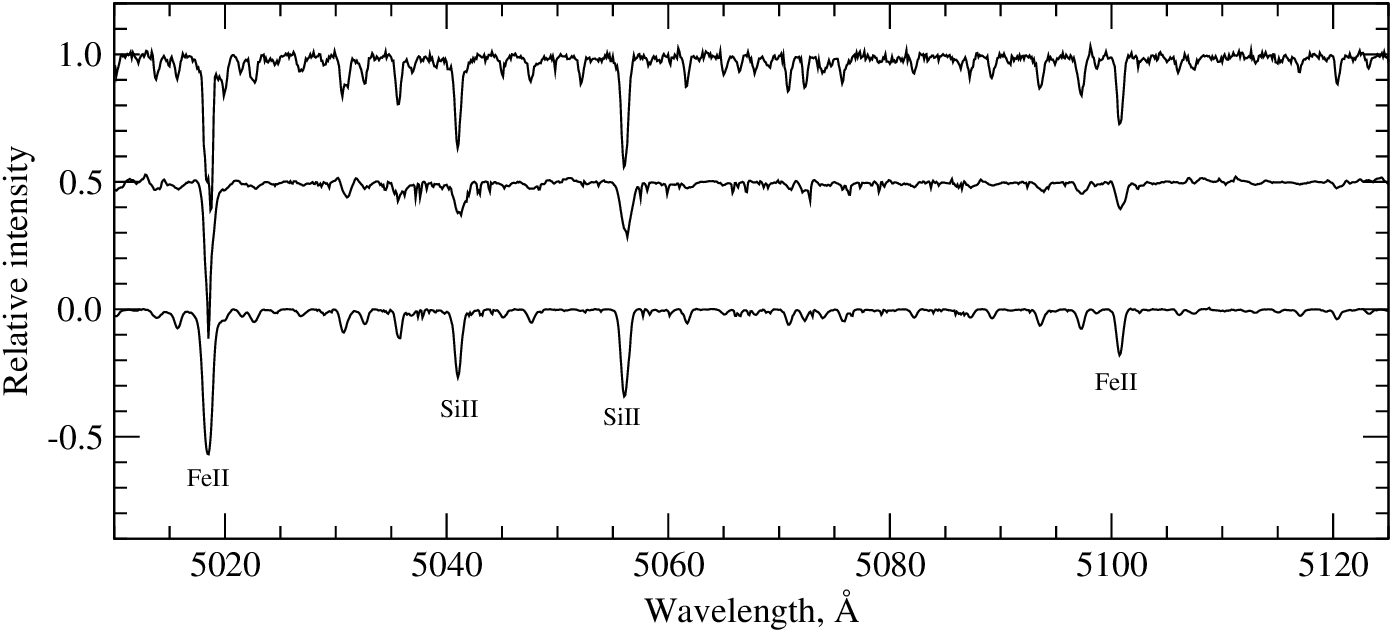}
}
\caption{} 
\label{fragment4}
\end{figure*}

\begin{figure*}
\vbox{
\includegraphics[angle=0,width=1.0\textwidth,bb=35 210 710 521,clip]{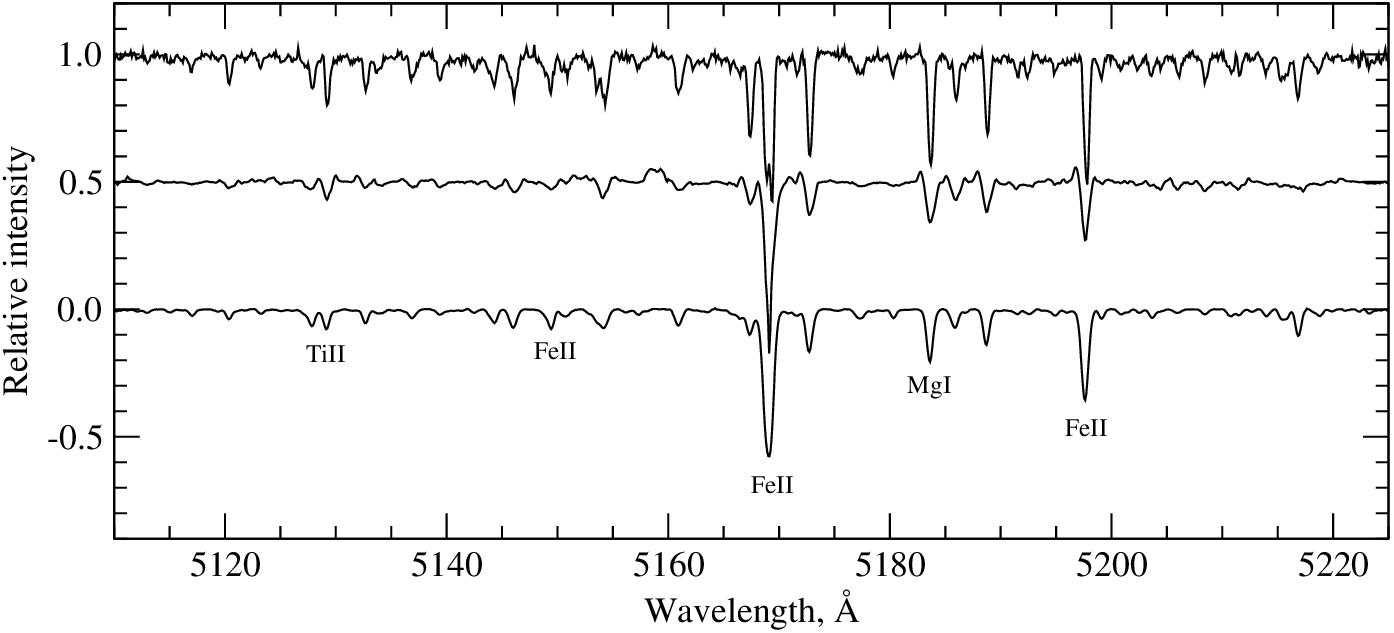}
\includegraphics[angle=0,width=1.0\textwidth,bb=35 210 710 521,clip]{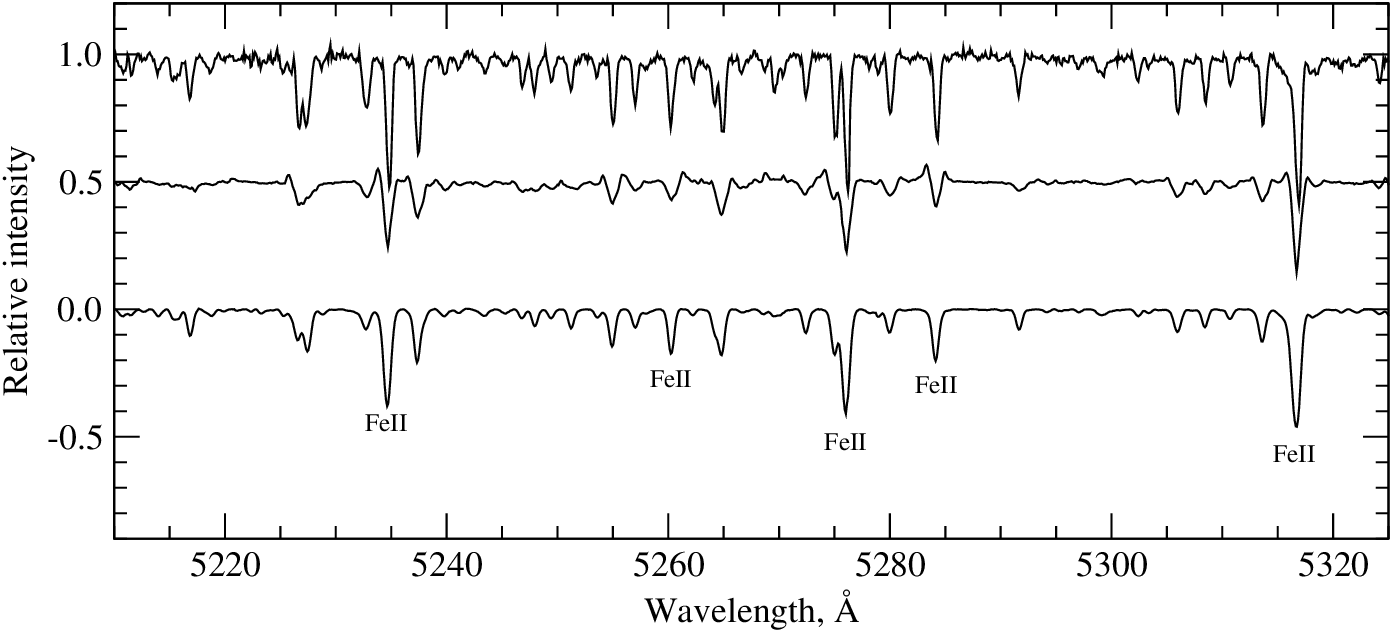}
\includegraphics[angle=0,width=1.0\textwidth,bb=35 210 710 521,clip]{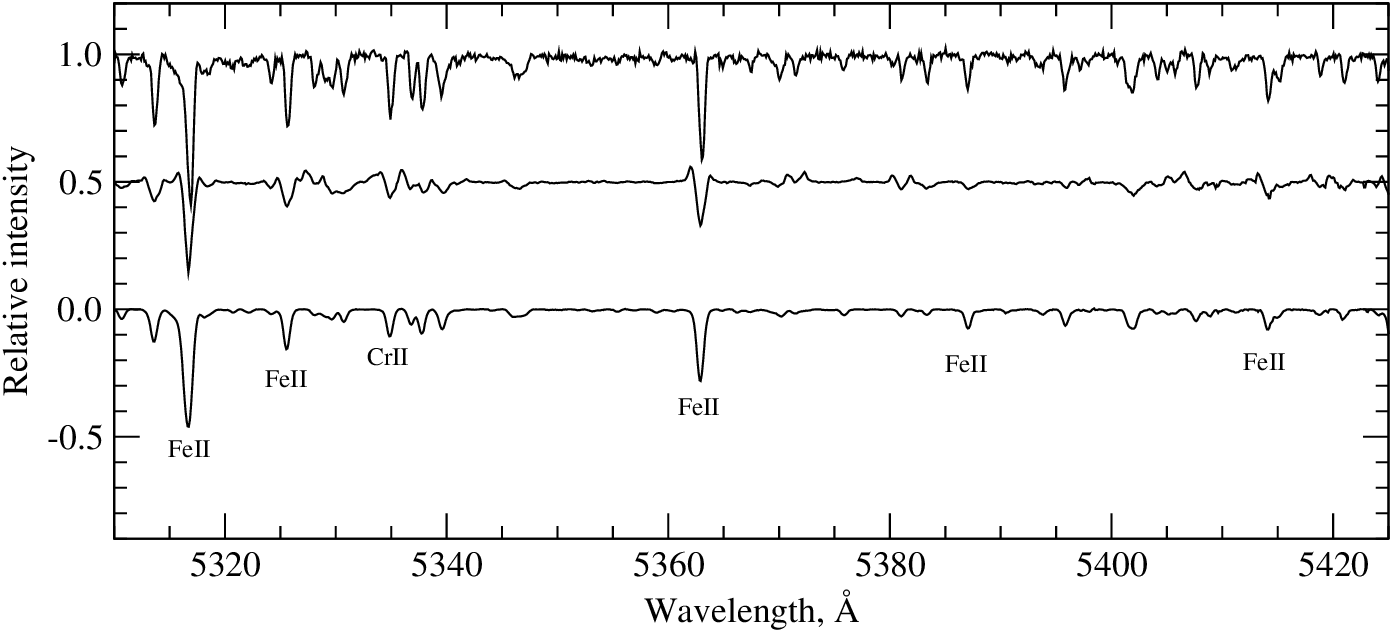}
}
\label{fragment5}
\end{figure*}

\begin{figure*}
\vbox{
\includegraphics[angle=0,width=1.0\textwidth,bb=35 210 710 521,clip]{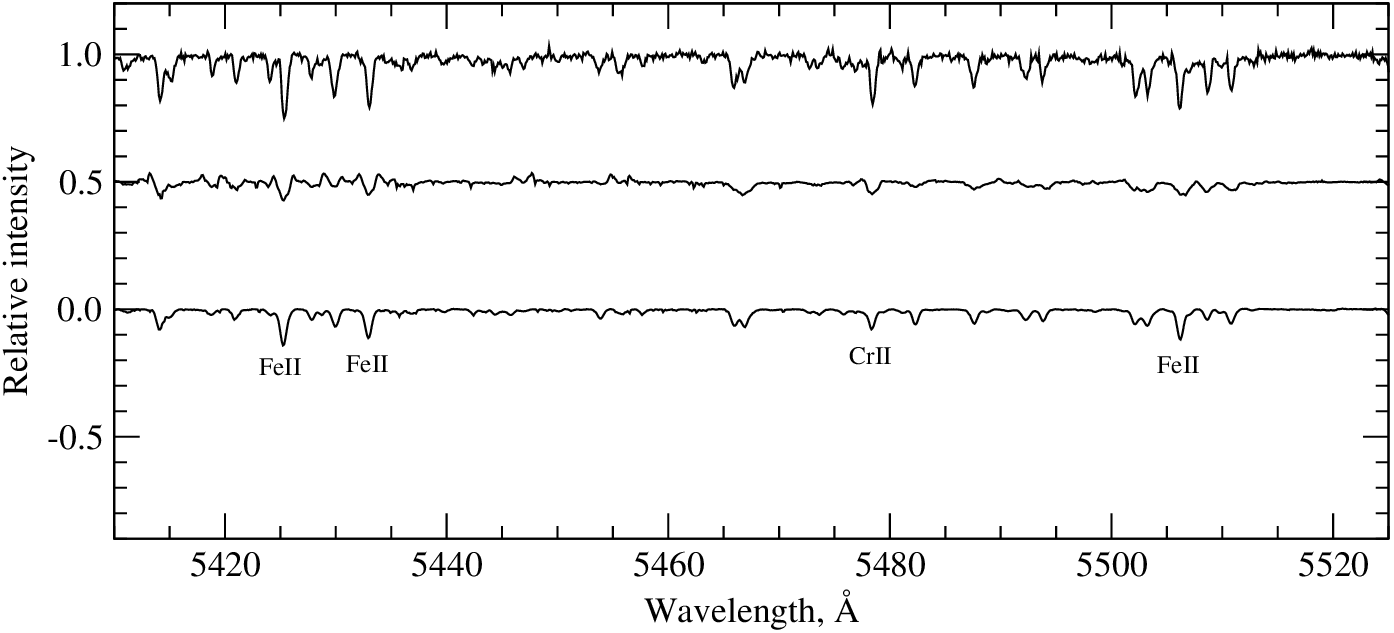}
\includegraphics[angle=0,width=1.0\textwidth,bb=35 210 710 521,clip]{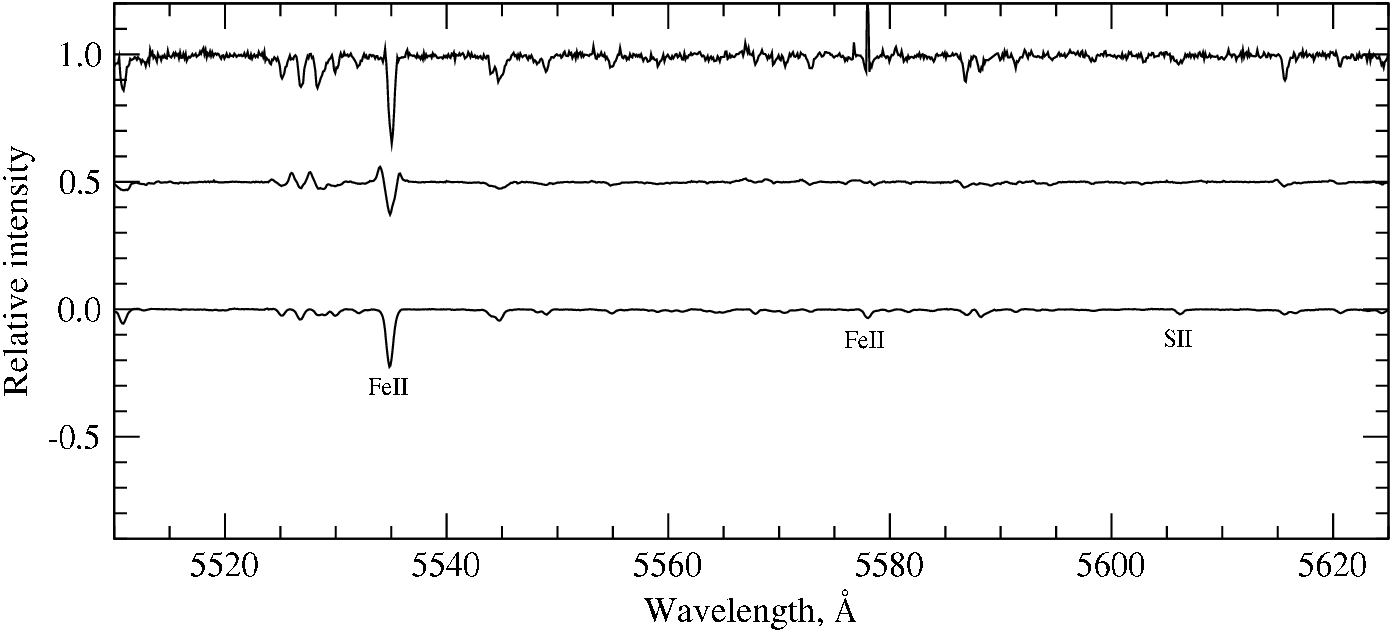}
\includegraphics[angle=0,width=1.0\textwidth,bb=35 210 710 521,clip]{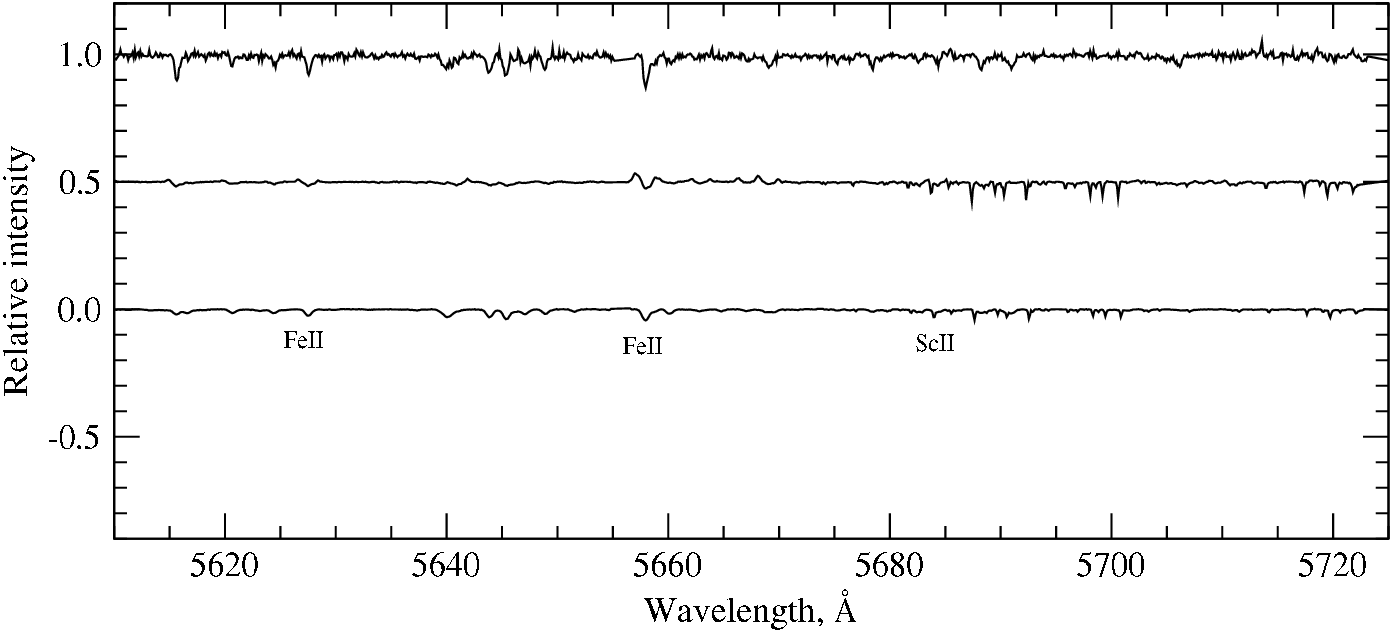}
}
\caption{}  
\label{fragment6}
\end{figure*}

\begin{figure*}
\vbox{
\includegraphics[angle=0,width=1.0\textwidth,bb=35 210 710 521,clip]{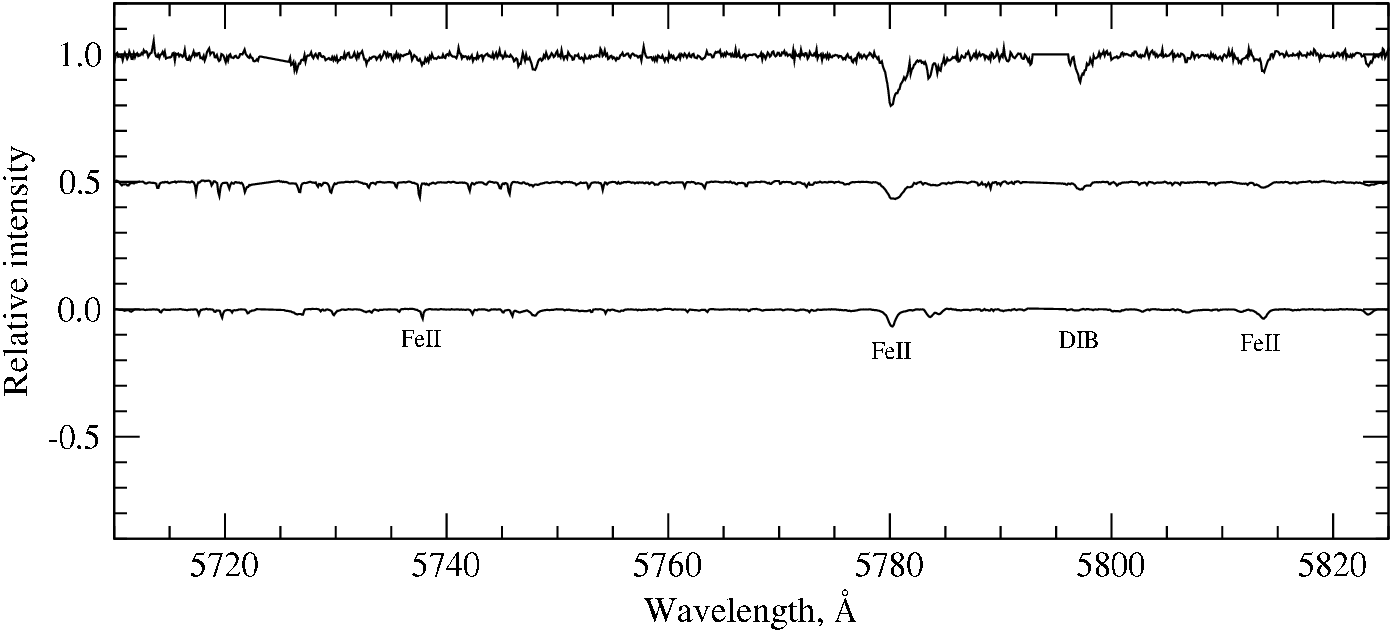}
\includegraphics[angle=0,width=1.0\textwidth,bb=35 210 710 521,clip]{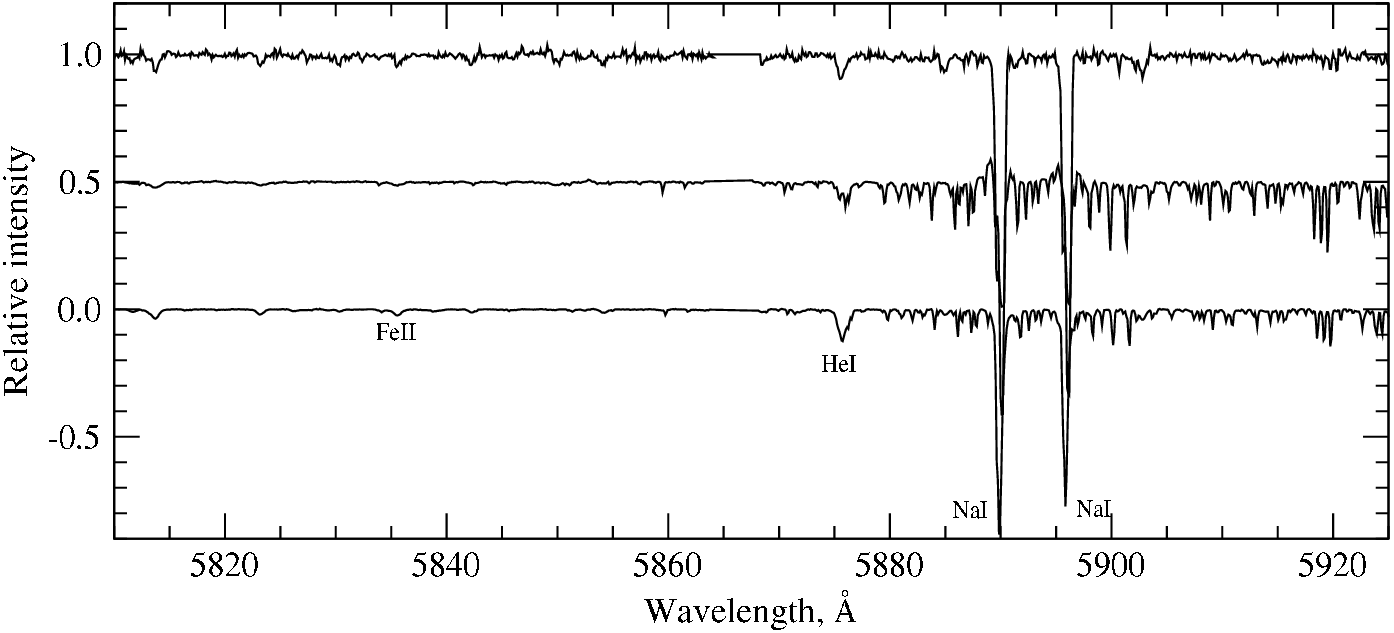}
\includegraphics[angle=0,width=1.0\textwidth,bb=35 210 710 521,clip]{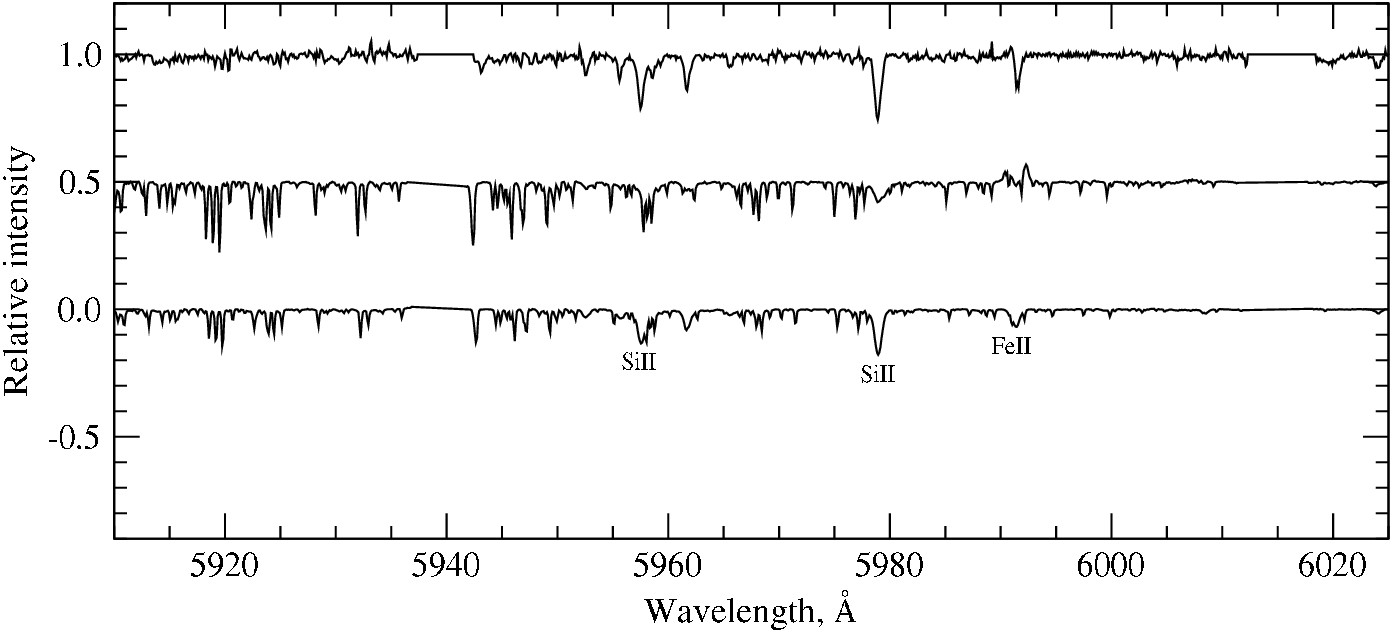}
}
\caption{ } 
\label{fragment7}
\end{figure*}

\begin{figure*}
\vbox{
\includegraphics[angle=0,width=1.0\textwidth,bb=35 210 710 521,clip]{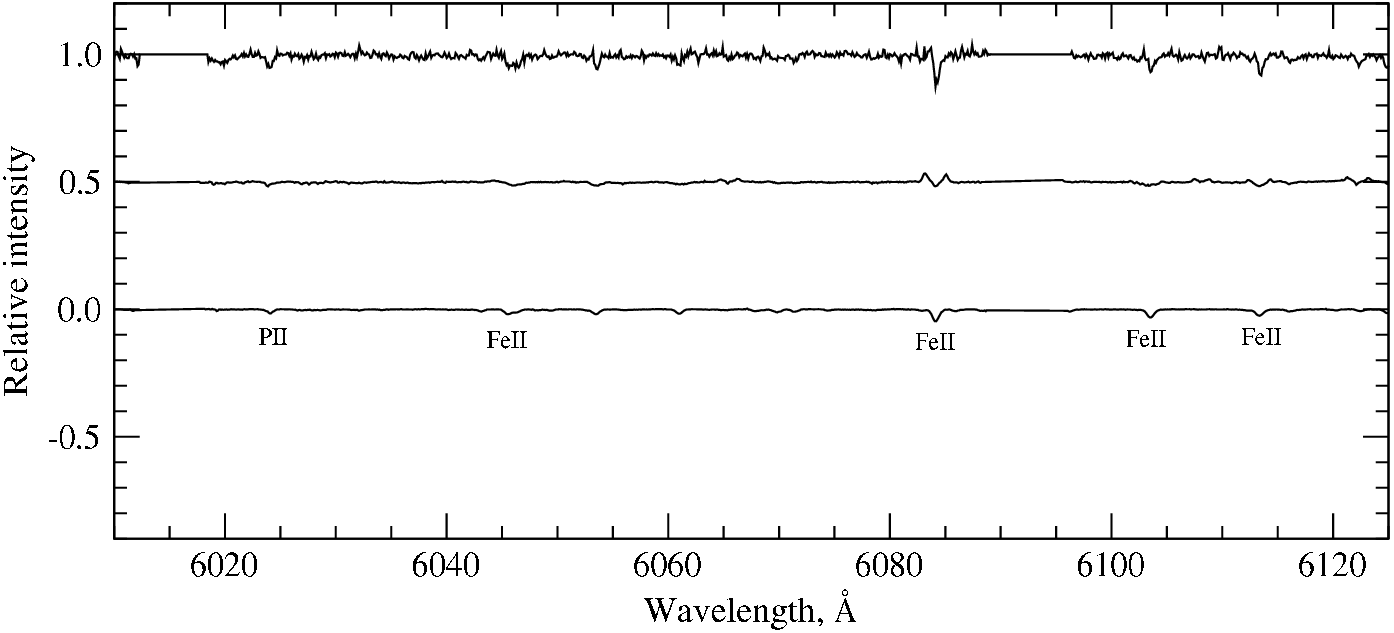}
\includegraphics[angle=0,width=1.0\textwidth,bb=35 210 710 521,clip]{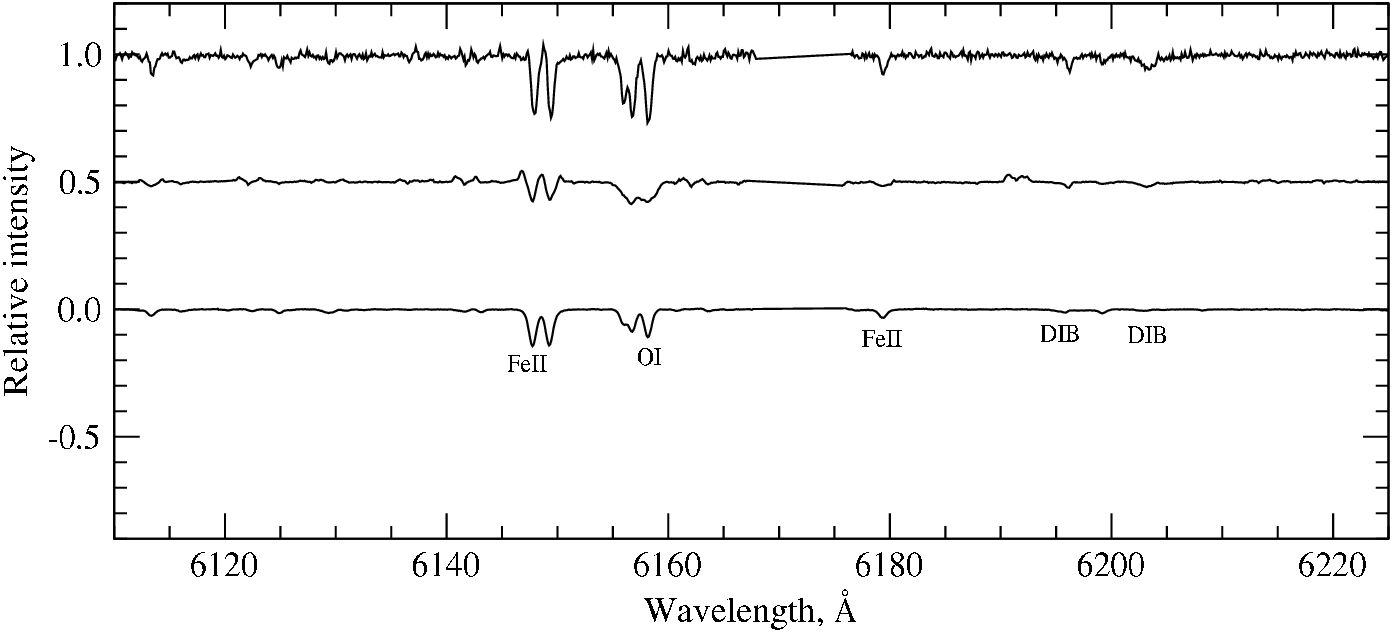}
\includegraphics[angle=0,width=1.0\textwidth,bb=35 210 710 521,clip]{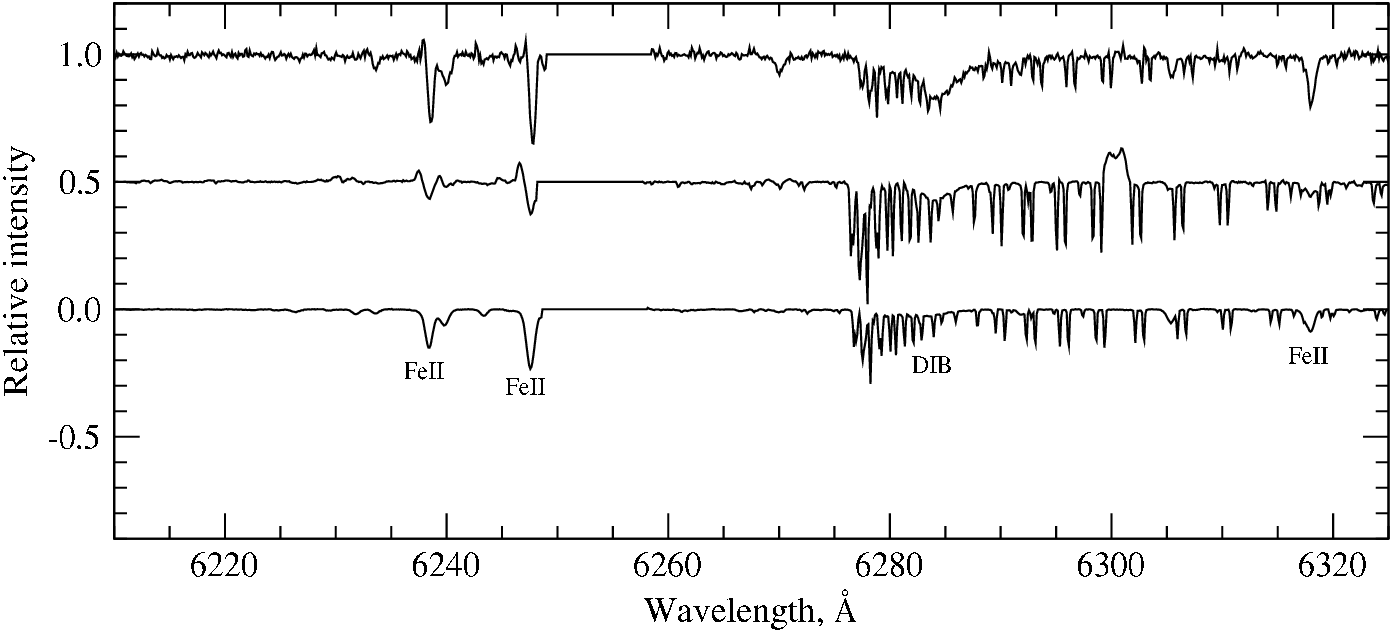}
}
\caption{The atlas fragments in the wavelength range of 6120--6520~\AA{}.}
\label{fragment8}
\end{figure*}

The spectra of the stars under study are presented in the atlas in the form of diagrams (see  Figs.~\ref{fragment1} --\ref{fragment9}).
These figures show the dependences of the residual intensity $r$ on the laboratory wavelength, they are 
arranged one below the other in the same order as the objects in Table~\ref{stars}. To reduce the observed spectra to
the laboratory wavelengths, we determined the radial velocities $V_r$ of the stars at the moment of their observation; the lines
of low and moderate intensity without any evident distortions were selected for this procedure. It should be emphasized that we have
not found differential shifts of these lines in the spectra of the studied stars. When changing the wavelengths to the laboratory
ones, this allowed us to apply the average values $V_r=-5.5$, +30.5, $-5.2$~km\,s$^{-1}$ for  BD\,$+48\degr1220$, 3\,Pup, and 
$\alpha$\,Cyg respectively.

For each star the $r$ ($\lambda$) values of individual echelle orders were combined into one array which was then divided into
equal fragments of 100~\AA{}. Each fragment of the spectral atlas shows the identification of several lines. For the identification
of the spectral details, we used the results from the papers~[\cite{3Pup,05040b}] and also the data from VALD (see~[\cite{VALD}] 
and the references to the previous publications regarding the earlier versions).

The results of line identification and measurement are shown in Table~\ref{Lines}. The first two columns give the name of 
the chemical element, the multiplet number, and the used laboratory wavelength according to~[\cite{VALD}]. The following 
columns of this table show, for each star, the central residual intensities of the absorptions $r$ and the heliocentric 
radial velocities $V_r$ in km\,s$^{-1}$ measured from the absorption cores of single lines or from distinct components 
of blends. We use central residual intensities instead of equivalent widths for ease of comparison of tabular and graphical 
data. The horizontal lines in the table separate single lines and blends. The entire fragments of the spectra and 
Table~\ref{Lines} can be found at  {\it http://www.sao.ru/hq/ssl/Deneb/atlas.html}. Let us further consider some 
features of the spectra of the stars under study.

\subsection{$\alpha$\,Cyg}

Atlas~[\cite{Albayrak2003}] presents the fragments of the $\alpha$\,Cyg spectrum  in a limited wavelength
region (3826--5212~\AA{}) with a resolution similar to ours but with a higher signal-to-noise ratio
($S/N=$ $300$--$1400$). All the absorptions from this atlas down to the faintest ones with residual intensities
$r<0.995$ are detectable in our spectrum. Residual intensities do not differ systematically. Chentsov et
al.~[\cite{UV-atlas}] published a spectral atlas for four stars in a wide range of wavelengths 3055--4520~\AA{},
including the UV radiation reaching the Earth's surface. In atlas~[\cite{UV-atlas}], the spectra of the well-studied
stars with similar temperatures but different luminosities ($\beta$\,Ori, $\alpha$\,Lyr, and $\alpha$\,Cyg) are compared with
the spectrum of a low metallicity A supergiant KS\,Per, the atmosphere of which is hydrogen-poor,
${\rm H/He}=3 \times 10^{-5}$.

The shortwave spectral region of $\alpha$\,Cyg studied in this paper is comparable with the longwave spectral region in
atlas~[\cite{UV-atlas}]. The spectra from both atlases were obtained with the same spectrograph NES with identical spectral
resolution, which makes it possible to investigate the spectrum of $\alpha$\,Cyg in a wide wavelength range \mbox {3055--6720}~\AA{}.

\begin{figure}[]
\includegraphics[angle=0,width=0.4\columnwidth,bb=40 40 340 530,clip]{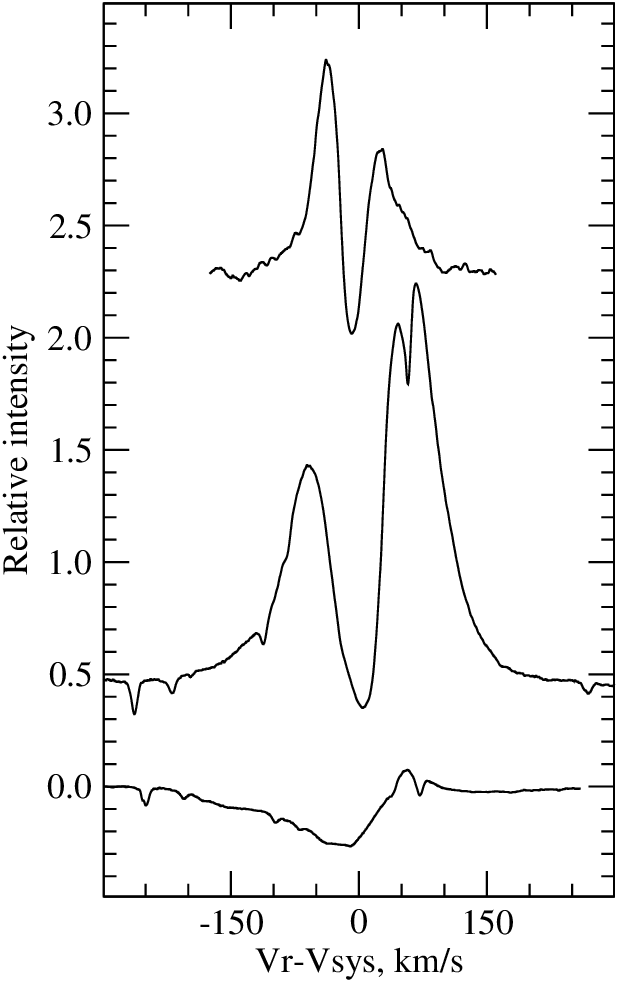}
\caption{Central parts of H$\alpha$ profiles in the spectra of the supergiants under study (top to bottom):
         BD\,$+48\degr1220$, 3\,Pup and $\alpha$\,Cyg. The H$\alpha$ profiles for 3\,Pup and BD\,$+48\degr1220$ are
         shifted upwards relative to the profile of the $\alpha$\,Cyg spectrum by 0.5 and 2.3 respectively.
         The difference of the heliocentric and systemic velocities $V_r-V_{\rm sys}$ in~km\,s$^{-1}$ is given 
         for each star on the horizontal axis. Telluric details are not removed.}
\label{Halpha}
\end{figure}

\subsection{BD\,$+48\degr1220$}

The optical spectrum of BD\,$+48\degr1220$ differs from the one of $\alpha$\,Cyg by the presence of numerous lines with asymmetric
profiles distorted by emission components. This difference is well illustrated by the profiles on the left and right sides of
Fig.~\ref{Profiles}. The H$\alpha$  line in the BD\,$+48\degr1220$ spectrum has a complex two-component emission
profile, for which time variability is typical, as follows from the comparison of the profile in Fig.~\ref{Halpha}
with the data from~[\cite{05040b}]. Let us note particularly that over the decade that passed since we obtained
the spectra used in~[\cite{05040b}], the ratio of the intensities of the shortwave and longwave emission components of
H$\alpha$ reversed. Variable emission components are also observed in the metal lines Si\,II, Fe\,I, Fe\,II. This variability is due
to the non-sphericity of the circumstellar shell, which is confirmed by investigations of the polarization of the light of
the star~[\cite{Partha2005}].

For a star with an effective temperature lower than 8000\,K, the presence of a strong absorption of neutral helium 
He\,I\,$\lambda$\,5876~\AA{} with an equivalent width of $W_{\lambda}(5875)=75$~m\AA{} is unexpected. The absorption
character of the line shows its photospheric nature; this is also proved by the fact that the radial velocity 
corresponding to the line position agrees with the velocity measured from other weak absorptions
(see Table~\ref{Lines}).

The presence of weak absorptions identified with well-known diffuse interstellar bands (DIBs) is also a peculiarity 
of the BD\,$+48\degr1220$ spectrum. They are listed in Table~\ref{Lines}, which also gives their equivalent widths 
and heliocentric velocities corresponding to the positions of these details. The average heliocentric velocity 
$V_r$ according to the interstellar bands agrees with the velocity measured from the main component of the Na~D lines:
$V_r({\rm DIBs})\approx -2$~km\,s$^{-1}$.

\subsection{3\,Pup}

\begin{figure*}
\vbox{
\includegraphics[angle=0,width=1.0\textwidth,bb=35 210 710 521,clip]{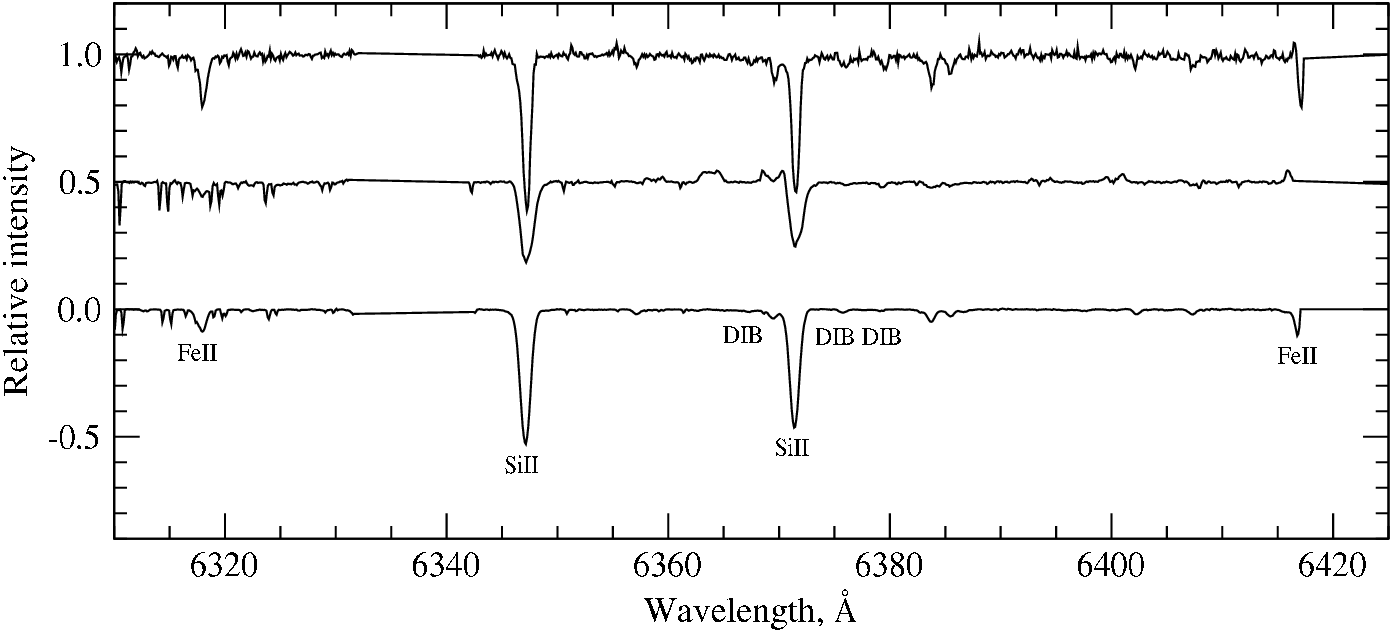}
\includegraphics[angle=0,width=1.0\textwidth,bb=35 210 710 521,clip]{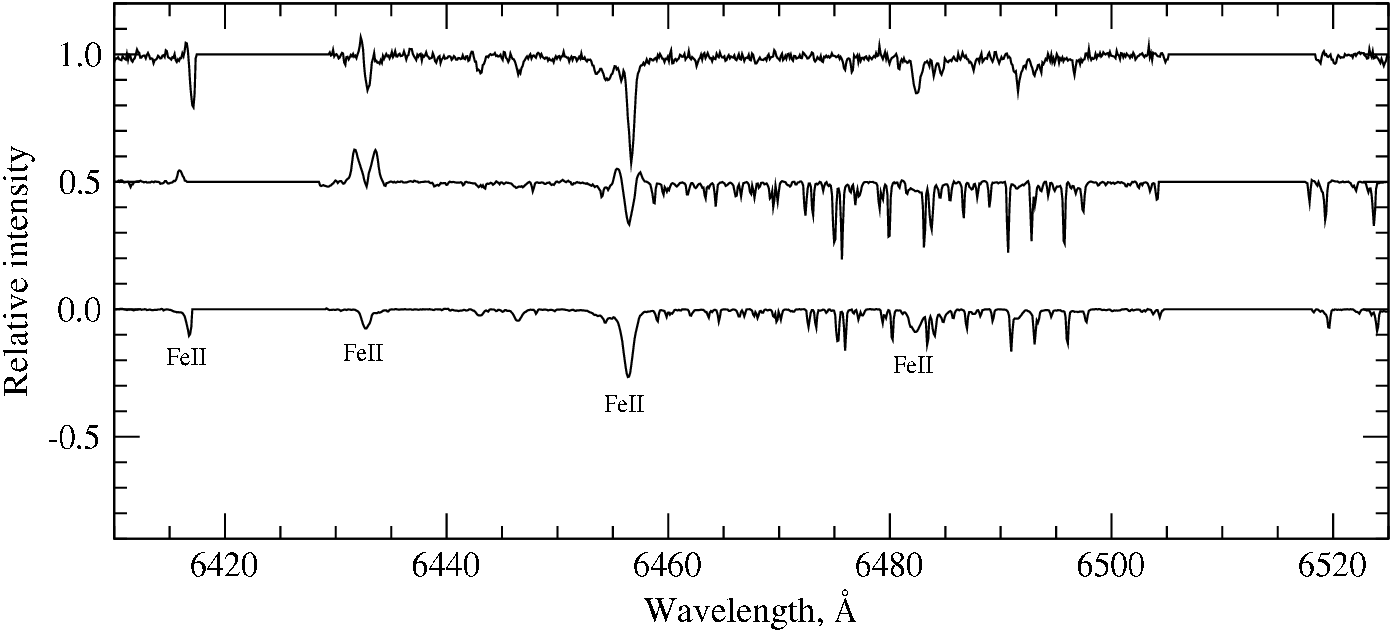}
\includegraphics[angle=0,width=1.0\textwidth,bb=35 210 710 521,clip]{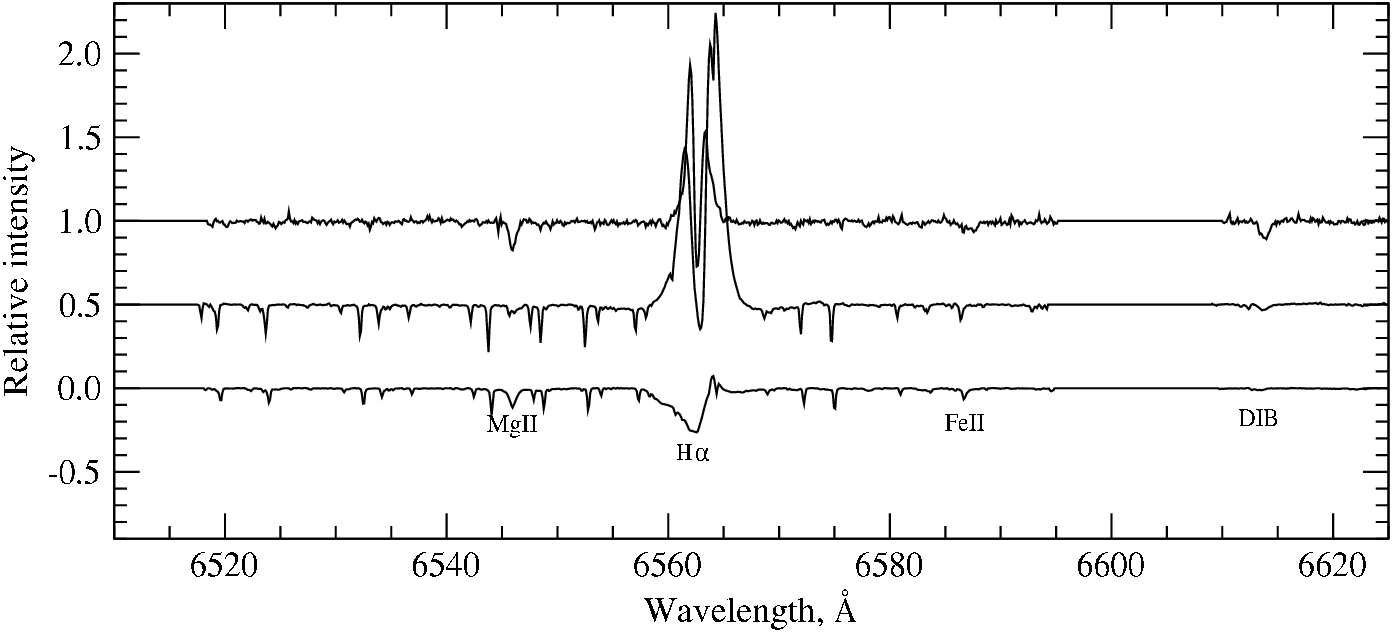}
}
\caption{Three longwave fragments from the atlas. The spectrum are  arranged top to bottom: 
         BD$+48\degr$\,1220, 3\,Pup, $\alpha$\,Cyg. The telluric details are not removed} 
\label{fragment9}
\end{figure*}

The characteristic feature of the 3\,Pup spectrum---the double-peak emission details of numerous line profiles (mainly of
Fe\,II ions)---is well reproduced in our atlas. In particular, such specific profiles with uplifted wings are presented in the
central part of Fig.~\ref{Profiles} and in Fig.~\ref{fragment9}. The  H$\alpha$ emission line in the 3\,Pup spectrum has 
a complex split profile with the red component prevailing (Fig.~\ref{Halpha}).

A useful comparison of the line profiles of different intensities in the spectra of 3\,Pup and $\alpha$\,Cyg was conducted
in~[\cite{3Pup}]. A comparison of the details shows that in the $\alpha$\,Cyg spectrum all the lines are symmetric and
differ in depth only. They are formed near the photosphere. Only the Mg\,II~4481~\AA{} line can be considered photospheric in
3\,Pup; as for the Fe\,II lines, here the shell contribution is apparent which gives the profiles a specific form: the wings are
uplifted by emissions, and the core is sharpened by a depression. The depression is noticeable even in the absorptions with the
residual intensity of $r\approx 0.2$, and it grows with the line strength. A comparison of the line profiles and their parameters
in the spectra of 3\,Pup and $\alpha$\,Cyg, obtained with one and the same spectral resolution, showed that the manifestation of the
shell in 3\,Pup is not limited to certain lines. It grows gradually with the increase of the intensity and wavelength of the
line. The central depths of absorptions in the spectra of 3\,Pup and $\alpha$\,Cyg are compared in Fig.\,1 of~[\cite{3Pup}].

The anomalies of the line profiles in the 3\,Pup spectrum naturally affect the measurements of radial velocities. The
$V_r(r)$ dependence derived from the measurements of lines in several spectra obtained on different dates is variable 
in time. However, for the forbidden emissions and the narrow shell cores of absorptions of the strong lines of the 
iron-group ions, the velocity variations in time are minimal, and the velocities derived from such lines are similar 
to each other. The average velocity for the mentioned lines can be assigned as the radial velocity of the whole system: 
$V_{\rm sys}=28.5\pm 0.5$~km\,s$^{-1}$~[\cite{3Pup}]. The average velocity $V_r$ measured from the interstellar DIBs is
$V_r=+28.55$~km\,s$^{-1}$.

\section{Conclusions}

Using high spectral resolution observations, we compiled a spectral atlas in the wavelength 
range of 3920--6720~\AA{} for three A supergiants with different evolutionary status: a massive
supergiant $\alpha$\,Cyg, a peculiar supergiant with a circumstellar disk 3\,Pup, and a post-AGB star 
BD\,$+48\degr1220$. A comparison of their spectra leads to the conclusion that the determination of 
the evolutionary status of supergiants in the galactic field is a sophisticated problem, as
the stars can be similar in spectral and luminosity classes but significantly different in age, mass, 
and observed on fundamentally different stages of evolution. One and the same region in the 
Hertzsprung--Russell diagram can be occupied by post-AGB stars evolving from the AGB stage to 
a planetary nebula, and massive supergiants evolving from the main sequence to the red supergiant stage. 
One spectacular example is the post-AGB star BD\,$+48\degr1220$ and the massive supergiant 3\,Pup with 
similar spectral features.

Practical experience with the spectra of stars of different types suggests that for definite conclusions 
on the evolutionary status of HLSs, the entire set of basic stellar parameters is required; 
foremost, the luminosity, the velocity field in the atmosphere, and the detailed chemical composition. 
High spectral resolution spectroscopy, which provides the data required for determining these parameters, 
is the main approach in stellar astrophysics. Spectropolarimetry and high spatial resolution observations 
also provide the necessary information.

\begin{acknowledgements}
The investigation was carried out with the support of the Russian Foundation for Basic Research (projects 12-07-00739a,
14-02-00291a). Observations with the 6-m BTA telescope are conducted with the financial support of the Ministry of
Education and Science of the Russian Federation (Agreement No.~14.619.21.0004, project identifier RFMEFI61914X0004). 
This research has made use of the SIMBAD and ADS databases.
\end{acknowledgements}

\clearpage
\newpage

{\footnotesize
\begin{longtable}{l|c @{{\vrule width 1pt}} c|c @{{\vrule width 1pt}} c|c @{{\vrule width 1pt}}  c|c @{{\vrule width 1pt}}  c|c }
\caption{Residual intensities $r$ and heliocentric radial velocities $V_r$~(km\,s$^{-1}$) for the spectral features in the
        high resolution spectra of $\alpha$\,Cyg, 3\,Pup, and BD\,$+48\degr1220$. The horizontal lines separate single lines 
        or unresolved blends. Uncertain measurements are marked with a colon} 
\label{Lines}
\\ \hline
\endfirsthead
\hline
Ident & $\lambda$ &\multicolumn{2}{c @{{\vrule width 1pt}}}{$\alpha$\,Cyg}
       &\multicolumn{2}{c@{{\vrule width 1pt}}}{\hspace{0.5cm}  3\,Pup \hspace{0.5cm} }&
      \multicolumn{2}{c@{{\vrule width 1pt}}}{ IRAS~05040  } \\
\cline{3-8}
    & & \qquad  {\it r} \qquad & $V_r$ (km/s) & \qquad {\it r} \qquad & $V_r$ (km/s) &\qquad {\it r} \qquad &
    $V_r$ (km/s)  \\
\hline
\endhead
\hline
\multicolumn{8}{r}{Table.\,1, be continued} \\
\endfoot
\endlastfoot
\hline
Ident & $\lambda$ &\multicolumn{2}{c@{{\vrule width 1pt}}}{$\alpha$\,Cyg}
    &\multicolumn{2}{c@{{\vrule width 1pt}}}{3\,Pup}
    &\multicolumn{2}{c@{{\vrule width 1pt}}}{IRAS~05040\qquad} \\
\cline{3-8} 
&&{\it r} & $V_r$ (km/s)&{\it r} &$V_r$ (km/s)&{\it r} & $V_r$ (km/s)\\ 
\hline
VII(10)&3916.41&0.93&$-$4.66&&&0.65$:$&$-$11.8$:$          \\
\hline                                                      
FeII(191)&3918.53&0.92&$-$4.11&&&0.77$:$&$-$12.4$:$              \\
CII&3918.97&&&&&&                                            \\
\hline                                                       
FeI(4)&3920.26&&&&&&                                         \\
FeII&3920.64&0.89&$-$8.58&&&0.68$:$&                              \\
CII(4)&3920.68&&&&&&                                         \\
\hline
\end{longtable}	     
}


\begin{thebibliography}{}

\bibitem{Massey} 1. P.~Massey, New Astron. Rev. \textbf{57},  14  (2013).

\bibitem{Hump2013} 2. R.M.~Humphreys, K.~Davidson, S.~Grammer,  et~al.,  \apj \textbf{773}, 46  (2013).

\bibitem{Kwok1993}  3. S.~Kwok,  \araa \textbf{31},  63  (1993).

\bibitem{IRC1} 4. V.G.~Klochkova,  E.L.~Chentsov, V.E.~Panchuk,  \mnras  \textbf{292}, 19 (1997).  
 
\bibitem{IRC2}  5. V.G.~Klochkova, M.V.~Yushkin, E.L.~Chentsov,  V.E.~Panchuk,  \arep  \textbf{46}, 139 (2002).

\bibitem{20572} 6. V.G.~Klochkova and E.L.~Chentsov, \ab  \textbf{63}, 112 (2008).

\bibitem{Ventura}  7. P.~Ventura, R.~Carini, and  F.D.~D'Antona,  \mnras \textbf{415}, 3865  (2011).

\bibitem{Lewis}  8. B.M.~Lewis,  \apj  \textbf{338}, 234 (1989).

\bibitem{Garcia}  9. P.~Garcia-Lario, A.~Manchado,  S.R.~Suso,  and  S.R.~Pottash, \aas \textbf{82},  497  (1990).

\bibitem{Arkhip}  10. V.P.~Arkhipova,  N.P.~Ikonnikova, R.I.~Noskova,  and  G.V.~Sokol, \alet \textbf{26}, 609 (2000).

\bibitem{Bright} 11. D.~Hoffleit and C.~Jaschek, {\it The Bright Star Catalogue}, 4th ed. (Yale University Observatory, New Haven, 1982).

\bibitem{Jager} 15. B.~Boer, C.~de~Jager, and  H.~Nieuwenhuijzen, \aaa \textbf{195}, 218 (1988).

\bibitem{05040b} 14.  V.G.~Klochkova, E.L.~Chentsov, N.S.~Tavolzhanskaya,  and   V.E.~Panchuk, \arep \textbf{51}, 642 (2007).

\bibitem{3Pup} 13. E.L.~Chentsov, V.G.~Klochkova,  and    A.S.~Miroshnichenko, \ab  \textbf{65}, 150 (2010).

\bibitem{Kaufer} 16. A.~Kaufer, O.~Stahl,  B.~Wolf, et~al., \aaa  \textbf{305}, 887 (1996).

\bibitem{Schiller} 12. F.~Schiller and    N.~Przybilla, \aaa \textbf{479}, 849 (2008).

\bibitem{Richard} 17. N.D.~Richardson,  N.D.~Morrison, E.E.~Kryukova,  and   S.J.~Adelman, \aj \textbf{141}, 17 (2011).

\bibitem{Saio} 18. H.~Saio,  C.~Georgy,  and   G.~Meynet,  \mnras \textbf{433}, 1246 (2013).

\bibitem{Jura} 19. M.~Jura  and  S.G.~Kleinmann, \apj \textbf{341}, 359 (1989).

\bibitem{Meilland} 20. A.~Meilland, S.~Kanaan, M.~Borges~Fernandes, et~al., \aaa  \textbf{512}, A73 (2010).

\bibitem{Trams} 21. N.R.~Trams, L.B.F.M.~Waters, H.J.G.L.~M.~Lamers,  et~ al., \aas \textbf{87}, 361 (1991).

\bibitem{Plets}  22. H.~Plets,  C.~Waelkens,  and   N.~R.~Trams, \aaa  \textbf{293}, 363 (1995).

\bibitem{Lamers} 23. H.J.G.L.M.~ Lamers,  F.J.~Zickgraf, D.~de~Winter, et~al., \aaa \textbf{340}, 117 (1998).

\bibitem{Oudm} 24. W.J.~de~Wit,  R.~Oudmaijer, and J.S.~Vink, Advances in Astronomy  \textbf{2014}, id.~270848 (2014).

\bibitem{Loden}  25. L.O.~Loden, The Messenger,  No.~68, 26 (1992).

\bibitem{Millour} 26. F.~Millour, A.~Meilland,  O.~Chesneau, et al., \aaa  \textbf{526}, A107 (2011).

\bibitem{Fujii}  27. T.~Fujii,  Y.~Nakada,  and  M.~Parthasarathy, \aaa  \textbf{385}, 884 (2002).

\bibitem{05040a} 28. V.G.~Klochkova,   E.L.~Chentsov, V.E.~Panchuk,   and  M.V.~Yushkin, IBVS,  No.~\,584,  (2004).

\bibitem{Rao} 29. S.S.~Rao, G.~Pandey,  D.L.~Lambert,  and   S.~Giridhar, \apj  \textbf{737}, L7 (2011).

\bibitem{nes} 30.  V.~Panchuk, V.~Klochkova, M.~Yushkin,  and I.~Najdenov, J.~Optical Technology  \textbf{76},  87 (2009).

\bibitem{slicer} 31.  V.~Panchuk, M.~Yushkin,  and   I.~Najdenov,  Preprint No.\,179, SAO RAS (Spec. Astrophys. Obs. RAS, 2003).

\bibitem{ECHELLE}  32. M.~Yushkin  and   V.~Klochkova,  Preprint No.\,206, (Special Astrophysical Observatory, Nizhnii Arkhyz, 2005).

\bibitem{gala} 33.  G.A.~Galazutdinov, Preprint No.\,92, SAO (Special Astrophysical Observatory, Nizhnii Arkhyz, 1992).

\bibitem{VALD} 34. F.~Kupka, N.~Piskunov,  T.A.~Ryabchikova,  et~al.,  \aas  \textbf{138}, 119 (1999).

\bibitem{Albayrak2003} 35. B.~Albayrak, A.F.~Gulliver, S.J.~Adelman, et~al., \aaa \textbf{400},  1043 (2003).

\bibitem{UV-atlas}  36. E.L.~Chentsov, V.G.~Klochkova, T.~Kipper, et.~al., \ab  \textbf{66},  466 (2011).

\bibitem{Partha2005} 37.  M.~Parthasarathy, S.~K.~Jain,  and   G.~Sarkar, \aj  \textbf{129}, 2451 (2005).

\end{thebibliography}
\end{document}